\documentclass{article}

\usepackage{PRIMEarxiv}

\usepackage[utf8]{inputenc} % allow utf-8 input
\usepackage[T1]{fontenc}    % use 8-bit T1 fonts
\usepackage{hyperref}       % hyperlinks
\usepackage{url}            % simple URL typesetting
\usepackage{booktabs}       % professional-quality tables
\usepackage{amsfonts}       % blackboard math symbols
\usepackage{nicefrac}       % compact symbols for 1/2, etc.
\usepackage{microtype}      % microtypography
\usepackage{lipsum}
\usepackage{fancyhdr}       % header
\usepackage{graphicx}       % graphics
\usepackage{amsmath}
\usepackage{amssymb}
\usepackage{amsbsy}
\usepackage{bigints}
\graphicspath{{media/}}     % organize your images and other figures under media/ folder

\newcommand*{\Resize}[2]{\resizebox{#1}{!}{$#2$}}

\title{Non-linear diffusion with stochastic resetting
}

\author{
  Przemys\l{}aw Che\l{}miniak \\
  Faculty of Physics \\
  Adam Mickiewicz University \\
  Uniwersytetu Pozna\'nskiego 2 \\
  61-614 Pozna\'n, Poland\\
  \texttt{geronimo@amu.edu.pl} \\
}

\begin{document}
\maketitle

\begin{abstract}
Resetting or restart, when applied to a stochastic process, usually brings its dynamics to
a time-independent stationary state. In turn, the optimal resetting rate makes the mean time
to reach a target to be the shortest one. These and other intriguing problems have been
intensively studied in the case of ordinary diffusive processes over the last decade. In this
paper we consider the influence of stochastic resetting on a diffusive motion modeled in terms
of the non-linear differential equation. The reason for its non-linearity is the power-law
dependence of the diffusion coefficient on the probability density function or, in another
context, the concentration of particles. We briefly outline this issue at first to prepare the
foundations for our further considerations. Then, we derive an exact formula for the mean
squared displacement and demonstrate how it attains the steady-state value under the influence
of exponential resetting. This mechanism brings also about that the spatial support of the
probability density function, which for the free non-linear diffusion is confined to the set
of a finite measure, tends to span the entire domain of real numbers. In addition, we explore
the first-passage properties for the non-linear diffusion intermittent by the exponential
resetting and find analytical expressions for the mean first-passage time and determine
numerically the optimal resetting rate which minimizes the mean time needed for a particle
to reach a pre-determined target. Finally, we test and confirm the universal property that
the relative fluctuation in the mean first-passage time of optimally restarted non-linear
diffusion is equal to unity.
\end{abstract}

\keywords{non-linear diffusion \and mean squared displacement \and probability density
\and stochastic resetting \and mean first-passage time
}

\section{Introduction}

Over the past few years a concept of stochastic resetting turned into a paradigm in such
diverse disciplines as statistical physics \cite{Magoni2020} and stochastic thermodynamics
\cite{Fuchs2016,Pal2017,Gupta2020}, biological physics \cite{Roldan2016,Budnar2019},
biochemistry \cite{Reuveni2014,Rotbart2015}, biology \cite{Pal2020} and computer science
\cite{Tong2008,Avrachenkov2013,Janson2012,Avrachenkov2018} drawing a significant attention
among scientific community. Resetting/restart is the procedure that stops and returns
a considered process at random instants of time to some pre-determined state, from which it
starts over again. Comprehensive studies performed in the last decade have confirmed that
this mechanism generates nontrivial effects on dynamical systems \cite{Evans2020}. For
instance, a system of which we certainly know that relaxes to the equilibrium
state, when subject to stochastic resetting, generically attains the time-independent
stationary state. Another property of resetting is to make a first-passage process finite
in time which might be difficult in some other circumstances.

A special type of processes for which the steady-state and the first-passage properties have
been intensively explored in the presence of resetting is diffusion. Intensified research on
this topic began with a publication of a seminal work on the canonical model of diffusion
through the agency of the exponential resetting \cite{Evans2011} (see also \cite{Majumdar2011}).
Since its original formulation this model has been gradually extended to include other
essential features of the  diffusive motion as well as the alternative variants of stochastic
resetting. Foremost examples comprise the models that consider diffusion confined in bounded
domains \cite{Christou2015,Chatterjee2018} and in external potentials with a constant
\cite{Pal2015,Roldan2017,Ray2020} and a space-dependent diffusivity \cite{Ray2020}, diffusion
in arbitrary spacial dimensions \cite{Evans2014}, under partial absorption
\cite{Whitehouse2013}, and in the presence of interactions \cite{Falcao2017,Basu2019}. Another
category of processes studied for stochastic resetting include anomalous diffusion
\cite{Mendez2019,Masoliver2019}, L\'evy flights \cite{Kusmierz2014,Kusmierz2015}, underdamped,
scaled and fractional Brownian motions \cite{Gupta2019,Bodrova2019,Sokolov2019,Wang2021},
random walk \cite{Campos2015,Majumdar2015,Montero2016,Mendez2016,Christophorov2021},
continuous-time random walk with and without drift \cite{Montero2013,Montero2017,Kusmierz2019},
telegraphic processes \cite{Masoliver2019a} and many others
\cite{Campos2016a,Chatterjee2019a,Meylahn2015}.

In the meantime, a few non-exponential resets and apart from them also more realistic
mechanisms of stochastic resetting have been proposed. In the former case, where the resetting
protocol is assumed to be instantaneous, one considered the diffusive systems with deterministic
\cite{Bhat2016}, non-static \cite{Santos2019}, scale-free \cite{Toyoizumi2019}, intermittent
\cite{Eule2016}, and non-Markovian \cite{Boyer2017} restarts as well as those with the space
and time-dependent resetting rates \cite{Kundu2016} and in the presence of the power-law
resetting time distributions \cite{Nagar2016}. However, bringing a material particle from one
place to another always takes some period of time in real situations. To make the resetting
processes more physical and practical form the experimental point of view
\cite{Roichman2020,Besga2020} one has needed to invent completely new solutions in comparison
with those discussed so far. Such theoretical ideas that have recently appeared involve
returns with a constant or space-dependent velocity
\cite{Pal2020a,Reuveni2019,Pal2019,Campos2019,Bodrova2020,Bodrova2020a}, a constant or
space-dependent acceleration \cite{Bodrova2020,Bodrova2020a}, resets with finite time
\cite{Evans2018} and under the influence of external traps such as confining potentials
\cite{Plata2021,Kundu2021}.

Besides studies on the stationary states of diffusive systems affected by resetting events
also their first passage properties has been extensively investigated at the same time
\cite{Reuveni2016b,Pal2017a,Chechkin2018}. This issue is particularly important
for the optimal search strategies
\cite{Benichou2005,Mallick2013,Catalan2009,Snider2012,Gudowska2015}
based on conceptually advanced algorithms dedicated to solve hard combinatorial problems
\cite{Montanari2002}. A typical quantitative measure of efficiency of a search process is
the mean-first passage time or, in other words, the time at which a searcher hits a target at
an unknown location for the first time on average \cite{Redner2001}. The main conclusion
arising from the studies conducted in this field is that restart renders the mean first-passage
time finite and minimal for the optimal resetting rate \cite{Evans2020}. It has been appeared
that the most effective search strategy is achieved for the deterministic resetting when the
time intervals between resets are of a constant value \cite{Kundu2016,Pal2017a,Chechkin2018}.
Aside from the optimal rate of resetting there exists a certain value of the exponent present
in the power-law distribution of waiting times between resets which for an ordinarily diffusing
particle minimizes its mean first-passage time from an initial to a given target position
\cite{Nagar2016}. The first-passage time problem under stochastic resetting has also been
analyzed in a presence of various external potentials \cite{Ahmad2019,Ray2020,Bier2017}, in
bounded domains \cite{Durang2019,Prasad2019} and under restart with branching
\cite{Eliazar2019}.

In all the aspects considered above the diffusive motion intermittent by stochastic resetting
was governed by the {\it linear} partial differential equations. As we know so far, the only
exception to this rule is the non-linear dynamical system such as chaotic Lorenz model
analyzed in the reference \cite{Hens2021}. However, this dynamics have nothing to do with a
diffusive motion. Thus, in the present paper we investigate the effect of exponential
resetting on both the steady-state and the first-passage properties of the {\it non-linear}
diffusion. Here, our intention is to consider the most fundamental aspects of this problem
and indicate those we intend to explore in the future.

In this paper we focus on a special variant of the non-linear diffusion equation in which
a diffusion coefficient depends on the probability density or concentration of particles
through the power-law relation with a constant exponent. We show that the mean squared
displacement and the probability density function resulting from the solution of the
non-linear differential equation have their counterparts in the form of appropriate
steady-state expressions as long as the non-linear diffusion proceeds under the influence
of exponential resetting. By contrast with the ordinary diffusion a domain of the probability
density function for the non-linear diffusion is restricted to a finite support, outside of
which this function disappears. However, the mechanism of exponential resetting makes the
support of the stationary probability density function extended to the entire domain of real
numbers. We will display this effect on the example of the exact as well as approximate
solutions of the non-linear diffusion equation. They have been obtained for peculiar values of
the power-law exponent that characterizes the dependence of the diffusion coefficient on the
probability density. In the second part of the paper we analyse the mean hitting time problem
and find the exact and approximate formulae for the mean first-passage time to the target
localized at the origin of the semi-infinite interval. Moreover, we also determine the optimal
resetting rate that minimizes this time and its dependence on the power-low exponent and the
distance from an initial position to the target. In addition, the relative standard deviation
associated with the mean first-passage time of the optimally restarted non-linear diffusion
at the constant rate is shown to be equal to unity.

The paper is organized as follows. In the next section we give a brief overview of the
non-linear diffusion equation with a power-law dependence of the diffusion coefficient on
the concentration/probability density. The informative content of this section is enough to be
used in subsequent parts of the paper. Sec.~3 is reserved to examine the basic stationary
properties of the non-linear diffusion under the exponential resetting from a viewpoint of the
mean squared displacement and the probability density function. In Sec.~4 we consider the
first-passage time properties of the non-linear diffusion interrupted by the exponential
resets. We summarize our results in Sec.~5.

\section{Non-linear diffusion equation}

In what follows we restrict our studies of the non-linear diffusion and its time course under
the influence of stochastic resetting to one dimension. Before we formulate a special type
of the equation describing this process let us firstly consider its more general form,
namely \cite{Debnath2012}
\begin{equation}
\frac{\partial}{\partial t}p(x,t)=\frac{\partial}{\partial x}\left(\mathcal{D}
\left[\,p(x,t)\right]\frac{\partial}{\partial x}p(x,t)\right).
\label{eq_1}
\end{equation}
By definition, this is a non-linear partial differential equation for a function $p(x,t)$,
which in a physical sense may stand for, depending on the context, the concentration of
diffusing particles, where $x$ is the distance from some initial position and $t$ is the
time, or the probability density function (PDF) of finding a diffusing particle in the
location $x$ at time $t$. In this paper we will consequently use the latter interpretation.
The reason for the non-linear nature of Eq.~(\ref{eq_1}) is a direct dependence of the
diffusivity $\mathcal{D}[p(x,t)]$ on the PDF through which it also depends on the variables
$x$ and $t$. Therefore, to specify the particular form of Eq.~(\ref{eq_1}) we have yet to
establish a specific relationship between the diffusion coefficient and the PDF. Due to many
interesting and practical applications that have attracted considerable attention within
a scientific community \cite{Fasano1986} we define this relation by the power-law function
\begin{equation}
\mathcal{D}=\mathcal{D}_{0}\left(\frac{p(x,t)}{p_{0}}\right)^{\sigma}.
\label{eq_2}
\end{equation}
In this expression $p_{0}$ denotes a constant reference value of a probability density, whereas
$\mathcal{D}_{0}$ is the diffusivity at that reference value. The power-law exponent $\sigma$
is a certain parameter. Only in the particular case for $\sigma\!=\!0$, Eq.~(\ref{eq_1})
converts into a linear diffusion equation with a diffusion constant $\mathcal{D}_{0}$. In turn,
if $\sigma\!>\!0$ then Eq.~(\ref{eq_1}) together with Eq.~(\ref{eq_2}) contribute to the
non-linear diffusion equation of the special type called the {\it porous medium equation}
\cite{Vazquez2007}. This differential equation has found many applications in a study of
such disparate transport phenomena as compressible gas flow through porous media
\cite{Barenblatt1990}, heat propagation occurring in plasma \cite{Berryman1978}, groundwater
flow in fluid mechanics \cite{Kochina1948}, population migration in biological environment
\cite{Gurtin1977,Murray2002}, the diffusion of grains in granular matter \cite{Christov2012}
and gravity-driven fluid flow in layered porous media \cite{Pritchard2001}, to name but a few
examples.

\begin{figure}[t]
\centering
\includegraphics[scale=0.3]{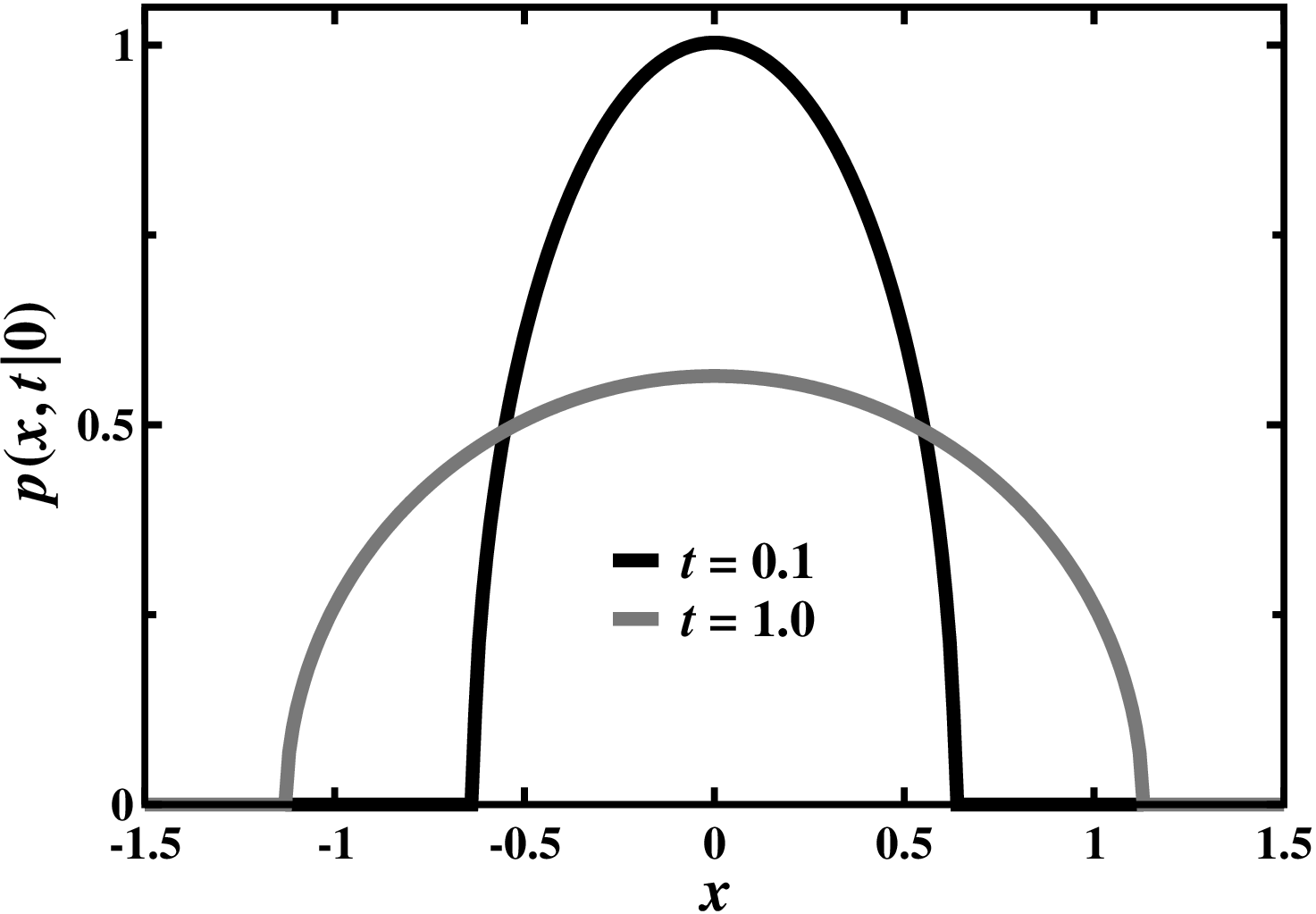}
\caption{Probability distribution function described by Eq.~(\ref{eq_6}) in two different
moments of time. Here we assumed the parameter $\sigma\!=\!2.0$ and the diffusion
coefficients $D\!=\!1.0$.}
\label{fig1}
\end{figure}

To give Eq.~(\ref{eq_1}) a more convenient form we now rewrite the diffusion coefficient
in Eq.~(\ref{eq_2}) so that $\mathcal{D}=Dp^{\sigma}(x,t)$. Here the parameter
$D=\mathcal{D}_{0}/p^{\sigma}_{0}$ is the generalized diffusion coefficient of the physical
dimension $[D]=\mathrm{L}^{\sigma+2}/\mathrm{T}$, where $\mathrm{L}$ and $\mathrm{T}$
are units of a length and a time, respectively. In consequence, the non-linear diffusion
equation is as follows:
\begin{equation}
\frac{\partial}{\partial t}p(t,x)=D\frac{\partial}{\partial x}\left(p^{\sigma}(x,t)
\frac{\partial}{\partial x}p(x,t)\right).
\label{eq_3}
\end{equation}
A commonly known procedure for solving this class of equations is offered by the method of
similarity solutions that utilizes an algebraic symmetry of a differential equation. In order
to find its solution we insert into Eq.~(\ref{eq_3}) a similarity transformation of the
algebraic form
\begin{equation}
p(x,t\!\mid\!x_{0})=\frac{1}{T(t)}F\left(\frac{x-x_{0}}{T(t)}\right)\equiv\frac{F(z)}{T(t)},
\;\;\text{with}\;\;z=\frac{x-x_{0}}{T(t)}
\label{eq_4}
\end{equation}
for the PDF of appearing a particle in $x$ at time $t$, if it was localized in the initial
position $x_{0}$ at time $t\!=\!0$. In this way we effectively reduce the original partial
differential equation for the non-linear diffusion to the system of two ordinary differential
equations for the separate functions $T(t)$ and $F(z)$ which are relatively easy to solve.
We omit detailed calculations here and refer the interested reader to \cite{Debnath2012},
where the discussed method is accessibly explained. Thus, as the final result we obtain
\begin{align}
p(x,t\!\mid\! x_{0})&=\frac{1}{T(t)}\left[a-\frac{b\sigma}{2D}\left(\frac{x-x_{0}}{T(t)}
\right)^{2}\right]^{\frac{1}{\sigma}},\;\;\text{with}\label{eq_5}\\
T(t)&=[b(\sigma+2)t]^{\frac{1}{\sigma+2}},\notag
\end{align}
where $a$ and $b$ are arbitrary integration constants. Their specific values can be
determined by adopting suitable boundary conditions. For example, if we set $a\!=\!1$,
with no substantiation for now, and impose the normalization condition
$\int_{-\infty}^{\infty}p(x,t\!\mid\!x_{0})dx=1$ on Eq.\ref{eq_5}, performing appropriate
integration with a help of the Euler beta function
$\mathrm{B}(\nu,\mu)=\frac{\Gamma(\nu)\Gamma(\mu)}{\Gamma(\nu+\mu)}=
2\int_{0}^{1}z^{2\nu-1}(1-z^{2})^{\mu-1}dz$ \cite{Gradshteyn2007}, we find the unknown $b$
and eventually a typical solution of Eq.(\ref{eq_3}) in the Zel'dovitch-Barenblatt-Pattle
algebraic form \cite{Zeldovich1958,Pattle1959,Barenblatt1972}
\begin{equation}
p(x,t\!\mid\!x_{0})=\frac{\mathcal{A}(\sigma)}{(D t)^{\frac{1}{\sigma+2}}}
\left[1-\mathcal{B}(\sigma)\frac{(x-x_{0})^{2}}{(D t)^{\frac{2}{\sigma+2}}}\right]
^{\frac{1}{\sigma}},
\label{eq_6}
\end{equation}
where the two $\sigma$-dependent coefficients in the above PDF are
\begin{equation}
\mathcal{A}(\sigma)=\left[\sqrt{\frac{\sigma}{2\pi (\sigma+2)}}\,
\frac{\Gamma\left(\frac{1}{\sigma}+\frac{3}{2}\right)}{
\Gamma\left(\frac{1}{\sigma}+1\right)}\right]^{\frac{2}{\sigma+2}}
\label{eq_7}
\end{equation}
and
\begin{equation}
\mathcal{B}(\sigma)=\sigma\left[\frac{1}{2(\sigma+2)}\right]^{\frac{2}{\sigma+2}}
\left[\frac{\sqrt{\pi}\,\Gamma\left(\frac{1}{\sigma}+1\right)}{\sqrt{\sigma}\,
\Gamma\left(\frac{1}{\sigma}+\frac{3}{2}\right)}\right]^{\frac{2\sigma}{\sigma+2}}.
\label{eq_8}
\end{equation}
The plot in Fig.~\ref{fig1} displaces profiles of the PDF in two different moments of time.
A supplementary comment is necessary at this point. The formulae given in Eqs.~(\ref{eq_5})
and (\ref{eq_6}) do not guarantee that the PDF for the non-linear diffusion is always
a real and positive function of $x$ as it should be by virtue of a very definition of the
probability density $p(x,t\!\mid\!x_{0})\geqslant0$. For this reason we need to assume the
additional requirement that the PDF can only be determined on the finite support
$\mid\!x-x_{0}\!\mid\leqslant\mathcal{B}^{-\frac{1}{2}}(Dt)^{\frac{1}{\sigma+2}}$.
Everywhere outside this interval the PDF vanishes and such a property was taken into
account when performing integration in the normalization condition to figure out the
parameter $b$.

We are still aware choosing a value of the parameter $a\!=\!1$ without any explanation which
may raise serious reservation. The argument behind such a choice is as follows. Supposing
$\sigma$ goes to zero and using the assertion that
$\lim_{z\to\infty}\frac{\Gamma(z+\alpha)}{\Gamma(z+\beta)}z^{\beta-\alpha}=1$
(see~\cite{Gradshteyn2007}) along with $\alpha=3/2$, $\beta=1$ and $z=1/\sigma$, we
obtain from Eqs.~(\ref{eq_7}) and (\ref{eq_8}) that $\mathcal{A}\!=\!1\!/\!\sqrt{4\pi}$
and $\mathcal{B}\!=\!\lim_{\sigma\to0}\sigma\!/\!4$. Simultaneously, expressing the
right-hand site of Eq.~(\ref{eq_6}) through the limit definition of the exponential
function $\mathrm{e}^{-z}=\lim_{n\to\infty}(1-\frac{z}{n})^{n}$ for $n\!=\!1/\sigma$,
we immediately retrieve the Gaussian distribution
\begin{equation}
p(x,t\!\mid\!x_{0})=\frac{1}{\sqrt{4\pi D t}}\exp\left[-\frac{(x-x_{0})^{2}}{4Dt}\right].
\label{eq_9}
\end{equation}
The above function is a fundamental solution of the linear partial differential equation
for a free diffusion given by Eq.~(\ref{eq_3}) whenever $\sigma\!=\!0$ with the initial
condition $p(x,0\mid x_{0})\!=\! \delta(x-x_{0})$. This result justifies our decision to
set $a\!=\!1$.

One of the crucial quantities characterising a diffusive motion is the mean squared
displacement (MSD). It is the most common measure of the deviation of the position $x$
of a diffusing particle with respect to some reference position $x_{0}$ over time. When
$x_{0}\!=\!0$ then MSD is defined as $\left<x^{2}(t)\right>\!=\!\int_{-\infty}^{\infty}
x^{2}p(x,t)\mathrm{d}x$. For the free linear diffusion in one dimension its MSD is
$\left<x^{2}(t)\right>\!=\!2Dt$, whereas in the case of the non-linear diffusion, for
which the PDF is given by Eq.~(\ref{eq_6}) we get
\begin{equation}
\left<x^{2}(t)\right>=D_{\sigma}t^{\frac{2}{\sigma+2}},
\label{eq_10}
\end{equation}
where the coefficient
\begin{equation}
D_{\sigma}=\frac{[2D(\sigma+2)]^{\frac{2}{\sigma+2}}}{3\sigma+2}\left[\frac{\sqrt{\sigma}
\,\Gamma\left(\frac{1}{\sigma}+\frac{3}{2}\right)}{\sqrt{\pi}\,\Gamma\left(\frac{1}{\sigma}
+1\right)}\right]^{\frac{2\sigma}{\sigma+2}}.
\label{eq_11}
\end{equation}
Again, if $\sigma\!\to\!0$ the above equations converge to the standard MSD for the
ordinary diffusion. Let us emphasize that contrary to this type of diffusion, the MSD
characterising the time course of the non-linear diffusion scales with the time according
to the power-law relation given by Eq.~(\ref{eq_10}). Moreover, if $\sigma\!>\!0$ then the
power-law exponent in this equation is less than unity which indicates that the non-linear
diffusion belongs to the class of the subdiffusive processes.

In the next part of the paper we explain how the presence of stochastic resetting affects
the MSD, Eq.~(\ref{eq_10}), and the PDF, Eq.~(\ref{eq_6}), that quantify the basic
properties of each diffusive motion including the non-linear diffusion as a special case.

\section{Non-linear diffusion with stochastic resetting}

A concept of stochastic resetting emerges from {\it renewal theory} being a part of a more
extensive theory of probability \cite{Cox1962}. Following this theory resetting is a mechanism
that interrupts a random process at random instants of time in such a way the process is stopped
and restarted anew from a given initial state. Here, we focus on the simplest version of this
mechanism assuming that the reset events happen instantaneously. Alternative scenarios of the
stochastic resetting regarding the non-linear diffusion will be analysed in successive papers.

Let us firstly consider a particle that executes {\it any} diffusive motion between reset
events. The renewal theory implicates that the PDF to find the particle at location $x$ at
time $t$, if it started from the initial position $x_{0}$ is
\begin{equation}
p_{r}(x,t\!\mid\!x_{0})=\int_{0}^{t}\Phi(t-\tau)p(x,t-\tau\!\mid\!x_{0})R(\tau)\mathrm{d}\tau,
\label{eq_12}
\end{equation}
where $p(x,t\!\mid\!x_{0})$ corresponds to the PDF for a free diffusion \cite{Eule2016}. We
supposed in the above equation without loss of generality that the particle is reset to
$x_{0}$ although this location can be arbitrary in a space. The function
\begin{equation}
\Phi(t)=\int_{t}^{\infty}\varphi(\tau)\mathrm{d}\tau=1-\int_{0}^{t}\varphi(\tau)\mathrm{d}\tau,
\label{eq_13}
\end{equation}
in Eq.~(\ref{eq_12}) denotes the probability (the survival probability) stating that no
resetting event occurs between two moments of time $0$ and $t$ and it is expressed by the time
integral of the distribution $\varphi(t)$ of waiting times between resets. This quantity is an
essential component of the integral equation
\begin{equation}
\varphi_{n}(t)=\int_{0}^{t}\varphi(t-\tau)\varphi_{n-1}(\tau)\mathrm{d}\tau,
\label{eq_14}
\end{equation}
that defines the probability $\varphi_{n}(t)$ that the $n$th reset event happens at time $t$
if the $(n-1$)th event occurred with the probability $\varphi_{n-1}(\tau)$ at the previous
moment of time $\tau$. The last quantity in Eq.~(\ref{eq_12}) is the probability
\begin{equation}
R(t)=\delta(t)+k(t),
\label{eq_15}
\end{equation}
of resetting events provided that the diffusion process starts with a reset event where the
$\delta$-distribution accounts for the initial condition and
\begin{equation}
k(t)=\sum\limits_{n=1}^{\infty}\varphi_{n}(t)
\label{eq_16}
\end{equation}
stands for the rate of resetting events. To determine this quantity we have to solve at first
the recursive relation in Eq.~(\ref{eq_14}). Due to the presence of the time convolution in this
equation we are capable of finding its solution in the Laplace domain. The result is as follows
\begin{equation}
\tilde{\varphi}_{n}(s)=\tilde{\varphi}^{n}(s),
\label{eq_17}
\end{equation}
where $\tilde{\varphi}(s)\!\equiv\!\int_{0}^{\infty}\varphi(s)\exp(-st)\mathrm{d}t$ is the
Laplace transform for which the convolution theorem has been applied. It states that the Laplace
transform of the convolution of two functions results in the product of their Laplace transforms.
Therefore, writing Eq.~(\ref{eq_16}) in the Laplace domain along with Eq.~(\ref{eq_16}) and
carrying out a summation of the geometric series we obtain the Laplace transform of the rate
\begin{equation}
\tilde{k}(s)=\frac{\tilde{\varphi}(s)}{1-\tilde{\varphi}(s)}.
\label{eq_18}
\end{equation}
Now, if we plugin Eq.~(\ref{eq_15}) into the primary Eq.~(\ref{eq_12}) then the PDF with the
stochastic resetting takes the following form
\begin{equation}
p_{r}(x,t\!\mid\!x_{0})=\Phi(t)p(x,t\!\mid\!x_{0})+\int_{0}^{t}k(\tau)\Phi(t-\tau)
p(x,t-\tau\!\mid\!x_{0})\mathrm{d}\tau.
\label{eq_19}
\end{equation}

For the particular case of the exponential (Poissonian) resetting, we focus in the present
paper on, the probability distribution of waiting times between resets, namely
\begin{equation}
\varphi(t)=r\mathrm{e}^{-rt},
\label{eq_20}
\end{equation}
corresponds to the exponential distribution where $r$ denotes a constant rate at which a
diffusive particle is reset to the initial position $x_{0}$. After making calculations of the
integral in Eq.~(\ref{eq_13}) we figure out the probability of no resets up to time $t$ is given
by $\Phi(t)\!=\!\mathrm{e}^{-rt}$. On the other hand, the Laplace transform of the function in
Eq.~(\ref{eq_20}) is $\tilde{\varphi}(s)\!=\!r/(s+r)$ and in consequence the rate of the
exponential resetting in the Laplace domain, Eq.~(\ref{eq_18}), takes the following form, i.e.
$\tilde{k}(s)\!=\!r/s$. Hence, performing the inverse Laplace transform we arrive at the
simple result that $k(t)\!=\!r$, whereas the PDF in Eq.~(\ref{eq_19}) for any diffusion with
the Poissonian resetting is now
\begin{equation}
p_{r}(x,t\!\mid\!x_{0})=\mathrm{e}^{-rt}p(x,t\!\mid\!x_{0})+r\int_{0}^{t}\mathrm{e}^{-r\tau}
p(x,\tau\!\mid\!x_{0})\mathrm{d}\tau.
\label{eq_21}
\end{equation}
The first component on the right-hand site refers to the trajectories of the process without
resets having occurred up to time $t$ with the probability $\mathrm{e}^{-rt}$, whereas the
second one corresponds to the phase of motion when the last resetting event took place at
$t\!-\!\tau$ with the probability $r\mathrm{e}^{-rt}$. The above equation constitutes the
starting point for the analysis of the MSD and the PDF for the non-linear diffusion under
the influence of the exponential resetting we will conduct in consecutive subsections.
\begin{figure}[t]
\centering
\includegraphics[scale=0.3]{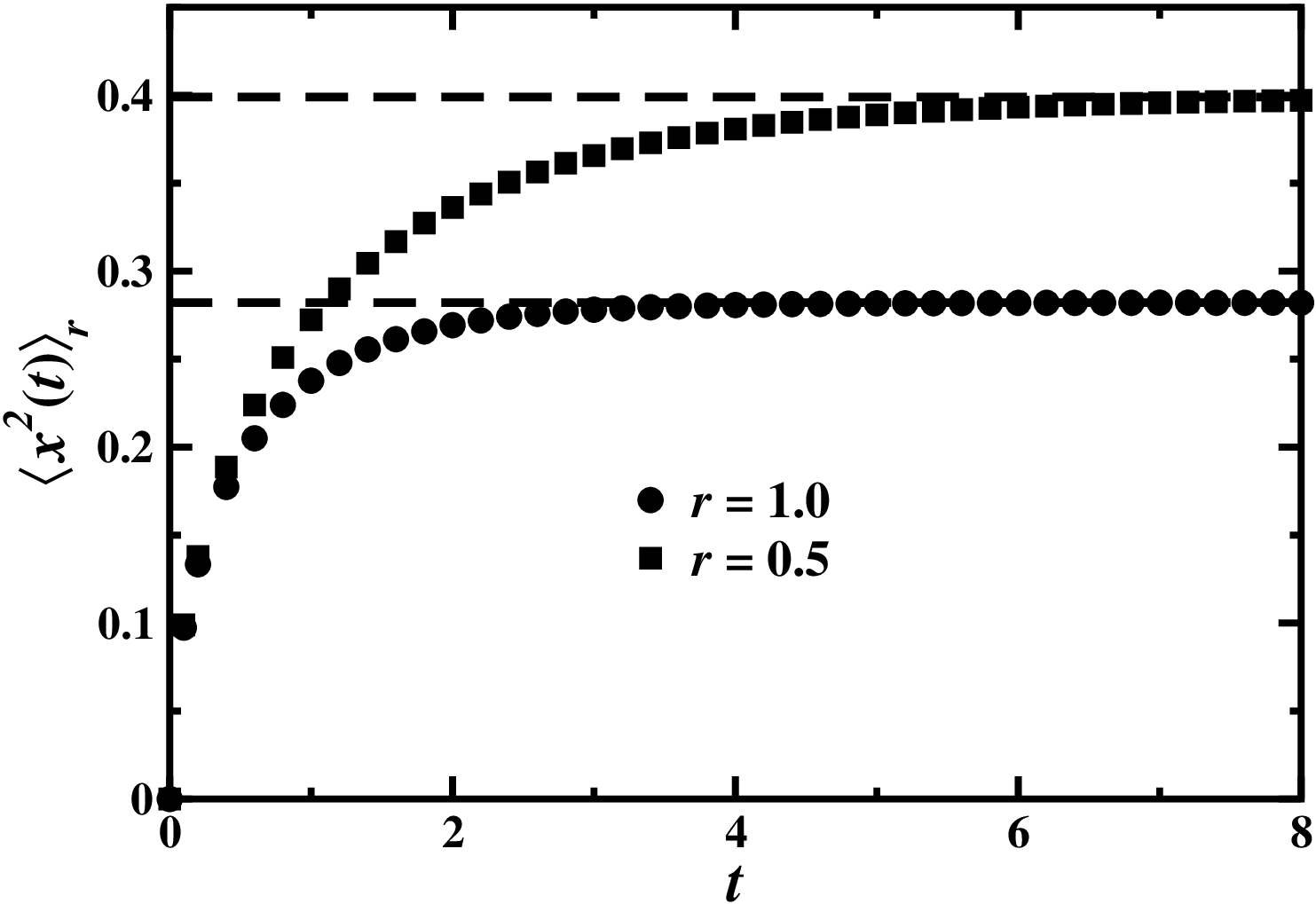}
\caption{Time course of the MSD $\left<x^{2}(t)\right>_{r}$ for the non-linear diffusion
with exponential resetting determined by two various resetting rates $r\!=\!0.5$ and $1.0$.
Striving for the MSD to attain stationary values (dashed lines) is an inherent feature of
diffusive motions and in general stochastic processes that proceed under the influence of
restart protocols.}
\label{fig2}
\end{figure}

\subsection{Mean squared displacement}

Multiplying Eq.~(\ref{eq_21}) by $x^{2}$ and performing integration over the variable $x$,
we obtain the equation for the MSD:
\begin{equation}
\left<x^{2}(t)\right>_{r}=\mathrm{e}^{-rt}\left<x^{2}(t)\right>+r\int_{0}^{t}
\mathrm{d}\tau\mathrm{e}^{-r\tau}\left<x^{2}(\tau)\right>.
\label{eq_22}
\end{equation}
The substitution of Eq.~(\ref{eq_10}) to the above formula leads after elementary calculations
to the exact result:
\begin{equation}
\left<x^{2}(t)\right>_{r}=D_{\sigma}t^{\frac{2}{\sigma+2}}\mathrm{e}^{-rt}+
D_{\sigma}r^{-\frac{2}{\sigma+2}}\gamma\left(\frac{\sigma+4}{\sigma+2},rt\right),
\label{eq_23}
\end{equation}
where $\gamma(\alpha,z)\!=\!\int_{0}^{z}u^{\alpha-1}\mathrm{e}^{-u}\mathrm{d}u$ is the lower
incomplete gamma function. Its asymptotic expansion for $z\!\to\!\infty$ is
\begin{equation}
\gamma(\alpha,z)\simeq\Gamma(\alpha)-z^{\alpha-1}\mathrm{e}^{-z}+(1-\alpha)
z^{\alpha-2}\mathrm{e}^{-z}
\label{eq_24}
\end{equation}
where $\Gamma(\alpha)\!=\!\int_{0}^{\infty}u^{\alpha-1}\mathrm{e}^{-u}\mathrm{d}u$ is the
complete gamma function which is represented here by the Euler integral of the second kind.
Using this expansion in Eq.~(\ref{eq_23}) we obtain the asymptotic expression for the MSD 
\begin{equation}
\left<x^{2}(t)\right>_{r}\simeq D_{\sigma}r^{-\frac{2}{\sigma+2}}
\left[\Gamma\left(\frac{\sigma+4}{\sigma+2}\right)-\frac{2}{\sigma+2}
(rt)^{-\frac{\sigma}{\sigma+2}}\mathrm{e}^{-rt}\right].
\label{eq_25}
\end{equation}
In turn, when $t\!\rightarrow\!\infty$ the MSD rapidly converges to the stationary state:
\begin{equation}
\left<x^{2}(t\!\to\!\infty)\right>_{r}=D_{\sigma}r^{-\frac{2}{\sigma+2}}
\Gamma\left(\frac{\sigma+4}{\sigma+2}\right)
\label{eq_26}
\end{equation}
In the special case for $\sigma\!\to\!0$ we reconstruct from Eq.~(\ref{eq_26}) the exact
result for the ordinary diffusion with the Poissonian resetting where it is known that
$\left<x^{2}(t)\right>_{r}\!=\!2D/r$. Fig.~\ref{fig2} displays the time course of the MSD as
described by Eq.~(\ref{eq_23}) for two different values of the resetting rates, $r\!=\!0.5$ and
$1.0$. We assumed here that the diffusion coefficient $D\!=\!1.0$, while the parameter
$\sigma\!=\!2.0$. The dashed lines indicate the stationary values of the MSD and their location
is established in accord with Eq.~(\ref{eq_26}).

\subsection{Probability density function}

From the very beginning we posit that the parameter $\sigma$ in Eq.~(\ref{eq_2}) is
assumed to be non-negative. By virtue of this requirement, the two $\sigma$-dependent
coefficients in Eq.~(\ref{eq_6}), namely $\mathcal{A}(\sigma)$ and $\mathcal{B}(\sigma)$
(see Eqs.~(\ref{eq_7}) and (\ref{eq_8}), respectively), are also non-negative and real.
In addition, the following condition $Dt\geqslant(\sqrt{\mathcal{B}(\sigma)}
\mid\!x-x_{0}\!\mid)^{\sigma+2}$ has to be satisfied to make the PDF in Eq.~(\ref{eq_6})
positive and real as the function depending on time. Moreover, the same condition will turn
out to be very crucial concerning integration performed over the time variable in
Eq.~(\ref{eq_21}).

To proceed, let us define the above restriction through the Heviside step function
$\Theta(z)$ which is equal to 1 if $z\geqslant0$ and 0 if $z<0$, and rewrite Eq.~(\ref{eq_6})
in the following form:
\begin{equation}
p(x,t\!\mid\!x_{0})=\Theta\!\left(t-D^{-1}(\sqrt{\mathcal{B}(\sigma)}\mid\!x-x_{0}\!\mid)^
{\sigma+2}\right)\frac{\mathcal{A}(\sigma)}{(D t)^{\frac{1}{\sigma+2}}}
\left[1-\mathcal{B}(\sigma)\frac{(x-x_{0})^{2}}{(D t)^{\frac{2}{\sigma+2}}}\right]
^{\frac{1}{\sigma}}.
\label{eq_27}
\end{equation}
If we now plugin this formula into Eq.~(\ref{eq_21}) and legally neglect its first
component supposing that $t\to\infty$, we obtain
\begin{equation}
p_{r}(x\!\mid\!x_{0})=r\int_{\xi(x,\sigma)}^{\infty}\frac{\mathcal{A}(\sigma)}
{(D t)^{\frac{1}{\sigma+2}}}\left[1-\left(\frac{\xi(x,\sigma)}{t}\right)^{\frac{2}
{\sigma+2}}\right]^{\frac{1}{\sigma}}\exp(-rt)\,\mathrm{d}t,
\label{eq_28}
\end{equation}
where the auxiliary function $\xi(x,\sigma)\!=\!D^{-1}(\sqrt{\mathcal{B}(\sigma)}
\mid\!x-x_{0}\!\mid)^{\sigma+2}$ allows us to simplify the formal notation. This expression
constitutes the starting point for our studies on the steady-state properties of the
non-linear diffusion interrupted by the exponential resetting. We explore this problem in
the present section whereas the issues related to the first-passage phenomena will be
considered in the subsequent section. Note, however, an explicit calculation of the integral
appearing in Eq.~(\ref{eq_28}) is no doubt a great challenge for arbitrary values of the
parameter $\sigma\!>\!0$. To avoid this difficult task we can make a use of computational
techniques utilizing either reasonable approximations or appropriate numerical procedures.
Nevertheless, there are a few exceptions from such solutions that concern the exact
results.

The case of $\sigma\!=\!\frac{1}{N}$, where $N$ represents any natural number, is rather
trivial and it will be not considered in the present paper. While the first exception we
begin with refers to the extreme situation when $\sigma\to0$. In this case
$\mathcal{A}\!=\!1/\sqrt{4\pi}$, $\xi(x,\sigma\to0)=\lim_{\sigma\to0}
\frac{\sigma}{4D}(x-x_{0})^{2}$ and hence the PDF in Eq.~(\ref{eq_28}) takes the following
form (see also the discussion in Sec.~2):
\begin{equation}
p_{r}(x\!\mid\!x_{0})=r\int_{0}^{\infty}\frac{1}{\sqrt{4\pi Dt}}
\exp\left[-\frac{(x-x_{0})^{2}}{4Dt}\right]\exp(-rt)\,\mathrm{d}t.
\label{eq_29}
\end{equation}
Using the more general integral $\int_{0}^{\infty}t^{\mu-1}\exp(-\alpha t^{-\gamma}-\beta 
t^{\gamma})dt\!=\!\frac{2}{\gamma}\left(\frac{\alpha}{\beta}\right)^{\mu/2\gamma}\mathrm{K}
_{\mu/\gamma}(2\sqrt{\alpha\beta})$ with $\alpha,\,\beta\!>\!0$ \cite{Gradshteyn2007}, where
$K_{\nu}(z)$ is the modified Bessel function of the second kind, which for $\nu\!=\!1/2$
corresponds to the exponential function, $K_{1/2}(z)\!=\!\sqrt{\frac{\pi}{2z}}
\mathrm{e}^{-z}$, we immediately retrieve the stationary probability distribution
\begin{equation}
p_{r}(x\!\mid\!x_{0})=\frac{1}{2}\sqrt{\frac{r}{D}}\exp\left(-\sqrt{\frac{r}{D}}\,
\lvert x-x_{0}\rvert\right),
\label{eq_30}
\end{equation}
for the ordinary diffusion under the influence of exponential resetting \cite{Evans2011}. Let
us clearly emphasize that when the parameter $\sigma$ goes to $0$ then simultaneously
the initially finite $x$-support of the sub-integral function, i.e. the PDF in
Eq.~(\ref{eq_27}), tends to span the entire real axis. In consequence, the $x$-domain of
the two-sided exponential PDF in Eq.~(\ref{eq_30}) extends from minus to plus infinity.
The same tendency will be observed for the $x$-support of the PDFs with $\sigma\!>\!0$
but, as we will show, a sole mechanism responsible for such a behavior is associated with
the exponential restart in a very long time limit.

A relatively simple instance when the integration in Eq.~(\ref{eq_28} can be exactly
performed refers to the parameter $\sigma\!=\!1.0$. In this case the coefficient
$\mathcal{A}\!=\!\left(\frac{3}{32}\right)^{1/3}$ and the lower limit of integration
$\xi(x,1)\!=\!\frac{2\mid x-x_{0}\mid^{3}}{9D}$, thus the final result is
\begin{equation}
p_{r}(x\!\mid\!x_{0})=\left(\frac{3r}{32D}\right)^{\frac{1}{3}}\left[\Gamma\!\left(
\frac{2}{3},\frac{2r}{9D}\!\mid\!x-x_{0}\!\mid^{3}\right)+\left(\frac{2r}{9D}\right)
^{\frac{2}{3}}(x-x_{0})^{2}\,\mathrm{Ei}\!\left(-\frac{2r}{9D}\mid\!x-x_{0}\!\mid^{3}
\right)\right],
\label{eq_31}
\end{equation}
where $\Gamma(\alpha,z)\!=\!\int_{z}^{\infty}u^{\alpha-1}\mathrm{e}^{-u}\mathrm{d}u$
corresponds to the upper incomplete gamma function and
$\mathrm{Ei}(-z)\!=\!-\int_{z}^{\infty}u^{-1}\mathrm{e}^{-u}\mathrm{d}u$  stands for the
exponential integral function. Note that for $x\!=\!x_{0}$ the above PDF approaches the
maximal value $p_{r}(x_{0})\!=\!\left(\frac{3r}{32D}\right)^{\frac{1}{3}}\Gamma\!\left(
\frac{2}{3}\right)$, because $\Gamma(z,0)\!=\!\Gamma(z)$ and $\lim_{z\to0}z^{2}\mathrm{Ei}
(-az^{3})\!=\!0$ which can be easy verified by virtue of the l'Hospital theorem. The same
result follows from a direct integration in Eq.~(\ref{eq_28}) with $\xi(0,1)\!=\!0$.

\begin{figure}[t]
\centering
\includegraphics[scale=0.3]{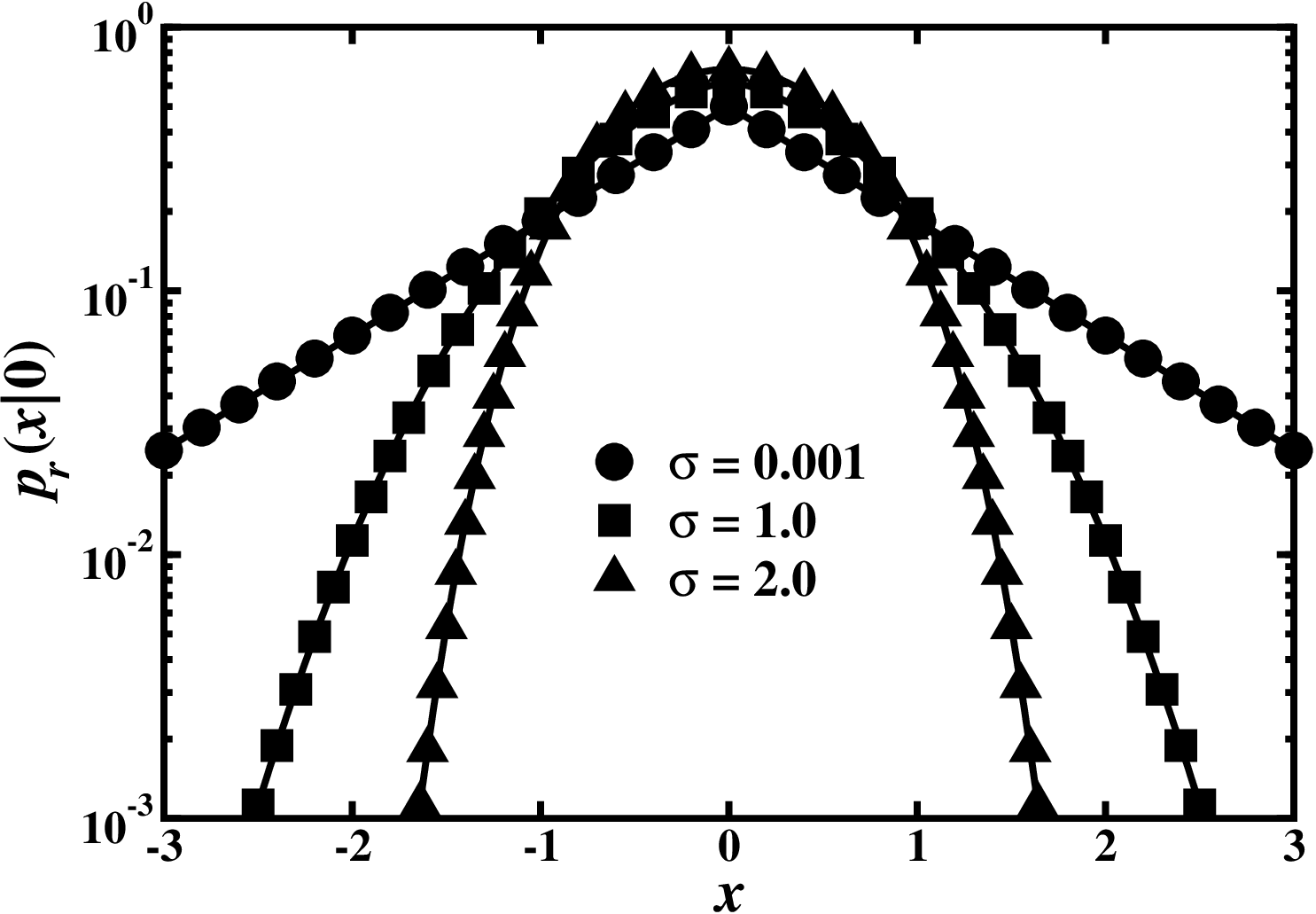}
\caption{The steady-state PDFs as a function of the position for different values of the
parameter $\sigma$. The points marked by circles, squares and triangles are outcomes received
from numerical integration performed in Eq.~(\ref{eq_28}). The solid lines represent the exact
analytical results given by Eqs.~(\ref{eq_30}), (\ref{eq_31}) and (\ref{eq_32}). The fixed
values of the diffusion coefficient $D\!=\!1.0$, the initial position $x_{0}\!=\!0$ of a
diffusing particle and the resetting rate $r\!=\!1.0$ have been assumed.}
\label{fig3}
\end{figure}

In contrast to the previous exceptions a slightly more complicated task of getting the
exact result from Eq.~(\ref{eq_28}) refers to the value of the parameter
$\sigma\!=\!2.0$. Here, we only show the final expression defining the PDF for the
non-linear diffusion intermittent by the exponential resetting which is as follows:
\begin{equation}
p_{r}(x\!\mid\!x_{0})=\frac{\pi r\!\mid\!x-x_{0}\!\mid^{3}}{8\sqrt{2}D}\exp\!\left(-
\frac{\pi^{2}r(x-x_{0})^{4}}{32D}\right)\left[\mathrm{K}_{\frac{3}{4}}\!\left(\frac{
\pi^{2}r(x-x_{0})^{4}}{32D}\right)-\mathrm{K}_{\frac{1}{4}}\!\left(\frac{\pi^{2}r(x-x_{0})
^{4}}{32D}\right)\right].
\label{eq_32}
\end{equation}
A detailed description of a derivation of this result is postponed into Appendix where we
also prove that $p_{r}(x_{0})\!=\!\left(\frac{r}{D}\right)^{1/4}\Gamma\left(\frac{3}{4}
\right) \!/\!\sqrt{\pi}$ for $x\!=\!x_{0}$. The plot displayed in Fig.~\ref{fig3}
shows the collection of PDFs for the peculiar values of the parameter $\sigma$ we have
considered until now. The points represented by circles, squares and triangles are outcomes
obtained from the numerical integration performed in Eq.~(\ref{eq_28}) and the solid lines
corresponds to the exact formulae given, respectively, by Eqs.~(\ref{eq_30}), (\ref{eq_31})
and (\ref{eq_32}). As expected, there is nothing special about the excellent consistency of
the numerical data with the analytical results. Nonetheless our intention at this point was
to demonstrate the correctness of the numerical method we will consequently use in this paper.
These three exact formulae for the PDFs are not only exceptional in the whole class of all
possible solutions that might be obtained from Eq.~(\ref{eq_28}) for arbitrary values of
the parameter $\sigma$. Together with the numerical tool we have at our disposal, they
constitute some kind of certainties for testing the approximate results. In the following
two such solutions called the algebraic and the exponential approximations are proposed.
The both arise from the fact that the PDF in Eq.~(\ref{eq_28}) must be by definition
non-negative and real quantity and this requirement is met if and only if
$\xi(x,\sigma)\!\leqslant\!t$. Hence, we can use as the first approximation the power-law
expansion $(1-z)^{\alpha}\!\simeq\!1-\alpha z$ with $\mid\!z\!\mid<1$ for the algebraic
approximation and the alternative relation $(1-z)^{\alpha}\!\simeq\!\exp(-\alpha z)$ for
the exponential approximation.

\begin{figure}[t]
\centering
\includegraphics[scale=0.3]{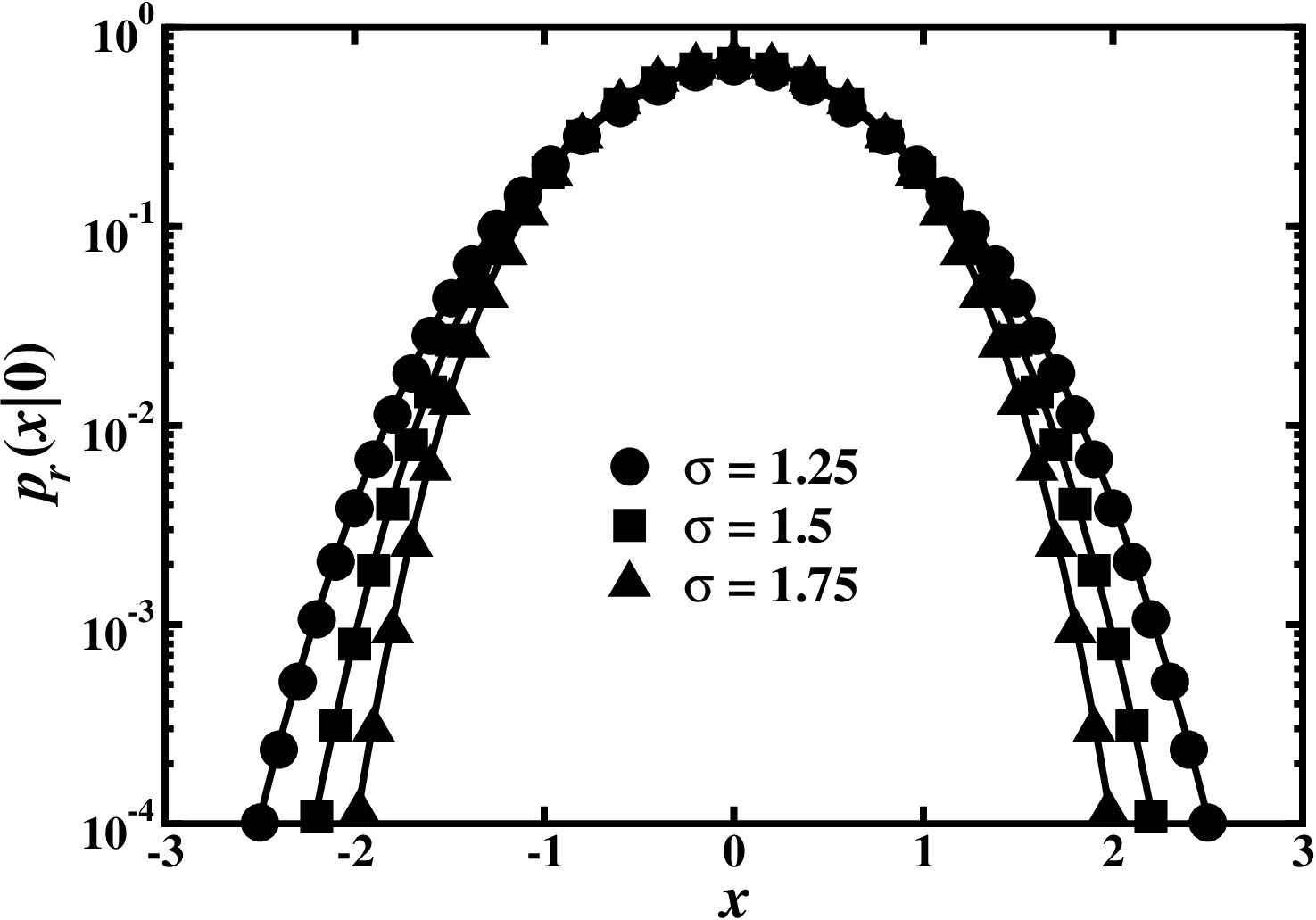}
\caption{The steady-state PDFs as a function of the position for different values of the
parameter $\sigma$ in the range between $1.0$ to $2.0$. The points marked by circles, squares
and triangles are outcomes received from the numerical integration performed in
Eq.~(\ref{eq_28}). The solid lines represent the approximate analytical result given by
Eqs.~(\ref{eq_35}). The fixed values of the diffusion coefficient $D\!=\!1.0$, the
initial position $x_{0}\!=\!0$ of a diffusing particle and the resetting rate $r\!=\!1.0$
have been assumed.}
\label{fig4}
\end{figure}

Let us consider at first the algebraic approximation. In this case the PDF given by
Eq.~(\ref{eq_28}) takes the following form:
\begin{equation}
p_{r}(x\!\mid\!x_{0})\simeq r\int_{\xi(x,\sigma)}^{\infty}\frac{\mathcal{A}(\sigma)}
{(D t)^{\frac{1}{\sigma+2}}}\left[1-\frac{1}{\sigma}\left(\frac{\xi(x,\sigma)}
{t}\right)^{\frac{2}{\sigma+2}}\right]\exp(-rt)\,\mathrm{d}t,
\label{eq_33}
\end{equation}
Now, we can proceed much the same as in a derivation of Eq.~(\ref{eq_31}). Decomposing the
above integral into two independent parts and calculating each of them separately, we
obtain after elementary operations that
\begin{equation}
p_{r}(x\!\mid\!x_{0})\simeq\mathcal{A}(\sigma)\left(\frac{r}{D}\right)
^{\frac{1}{\sigma+2}}\left[\Gamma\left(\frac{\sigma+1}{\sigma+2},r\xi(x,\sigma)\right)-
\frac{1}{\sigma}(r\xi(x,\sigma))^{\frac{2}{\sigma+2}}
\Gamma\left(\frac{\sigma-1}{\sigma+2},r\xi(x,\sigma)\right)\right]
\label{eq_34}
\end{equation}
In the last step we can extract the first incomplete gamma function in front of the
square bracket recasting the above formula as the power-law equation and eventually insert
in it the explicit form of the auxiliary function $\xi(x,\sigma)$. Therefore, the final
result is
\begin{equation}
p_{r}(x\!\mid\!x_{0})\simeq\mathcal{A}(\sigma)\left(\frac{r}{D}\right)
^{\frac{1}{\sigma+2}}\Gamma\left(\frac{\sigma+1}{\sigma+2},\frac{r}{D}\zeta(x,\sigma)\right)
\left[1-\left(\frac{r}{D}\zeta(x,\sigma)\right)^{\frac{2}{\sigma+2}}
\frac{\Gamma\left(\frac{\sigma-1}{\sigma+2},\frac{r}{D}\zeta(x,\sigma)\right)}
{\Gamma\left(\frac{\sigma+1}{\sigma+2},\frac{r}{D}\zeta(x,\sigma)\right)}\right]
^{\frac{1}{\sigma}},
\label{eq_35}
\end{equation}
where we have defined the new function $\zeta(x,\sigma)\!=\!(\sqrt{\mathcal{B}(\sigma)}
\mid\!x-x_{0}\!\mid)^{\sigma+2}$ for brevity of the notation. It is clear that if
$x\!=\!x_{0}$ then $\zeta(x_{0},\sigma)\!=\!0$ and hence the incomplete gamma function
$\Gamma(z,0)\!=\!\Gamma(z)$. We infer from these two properties that
\begin{equation}
p_{r}(x_{0})\simeq\mathcal{A}(\sigma)\left(\frac{r}{D}\right)
^{\frac{1}{\sigma+2}}\Gamma\left(\frac{\sigma+1}{\sigma+2}\right).
\label{eq_36}
\end{equation}
The plot of the PDF given by Eq.~(\ref{eq_35}) for disparate values of the parameter $\sigma$
falling in the range between $1.0$ and $2.0$ is shown in Fig.~\ref{fig4}. Here, we present the
compatibility of the algebraic approximation of the PDF (solid lines) with data represented
by points obtained from the numerical integration conducted in Eq.~(\ref{eq_28}) for the
non-linear diffusion with the resetting rate $r\!=\!1.0$. In addition, we have also verified
that the graph of the function plotted according to the approximate formula in Eq.~(\ref{eq_35})
deviates from the numerical results whenever the parameter $\sigma\!<\!1.0$. This fact gives
rise to a necessity of finding an appropriate expression for the PDF which could be also
applicable in the range of the parameter $\sigma$ between $0$ and $1.0$. We will not resolve
this problem in the present paper. Apart from this we conclude that the function in Eq.~(\ref{eq_35}
cannot be converted to the canonical form given by Eq.~(\ref{eq_30}) for $\sigma\to0$.

Let us now turn to the second case of the exponential approximation according to which
the term enclosed within square bracket in Eq.~(\ref{eq_6}) can be replaced by the
exponential function
\begin{equation}
p(x,t\!\mid\!x_{0})\simeq\frac{\mathcal{A}(\sigma)}{(Dt)^{\frac{1}{\sigma+2}}}\exp
\left[-\frac{\mathcal{B}(\sigma)(x-x_{0})^{2}}{\sigma(D t)^{\frac{2}{\sigma+2}}}\right].
\label{eq_37}
\end{equation}
The substitution of the above formula into Eq.~(\ref{eq_21}) with neglecting of the first
component in this equation leads to
\begin{equation}
p_{r}(x,t\!\mid\!x_{0})\simeq r\mathcal{A}(\sigma)\int_{0}^{t}\mathrm{d}\tau\,
(D\tau)^{-\frac{1}{\sigma+2}}\exp[-\theta(\tau)],
\label{eq_38}
\end{equation}
where we defined the function $\theta(\tau)\!=\!r\tau+\frac{\mathcal{B}(\sigma)(x-x_{0})
^{2}}{\sigma(D\tau)^{\frac{2}{\sigma+2}}}$ that poses a single minimum at
\begin{equation}
\tau_{0}=\frac{1}{D^{\frac{2}{\sigma+4}}}\left[\frac{2\mathcal{B}(\sigma)(x-x_{0})^{2}}
{r\sigma(\sigma+2)}\right]^{\frac{\sigma+2}{\sigma+4}}.
\label{eq_39}
\end{equation}
Because of the condition $0\ll\tau_{0}\ll t$ that holds for very long times (in practice,
for $t\!\to\!\infty$), we can apply to Eq.~(\ref{eq_38}) the Laplace saddle-point method
\cite{Wong2001} which after elementary calculations gives the PDF of the following form:
\begin{align}
p_{r}(x,t\!\mid\!x_{0})&\simeq r\mathcal{A}(\sigma)(D\tau_{0})^{-\frac{1}{\sigma+2}}
\exp[-\theta(\tau_{0})]\int_{-\infty}^{\infty}\mathrm{d}\tau\exp\left[-\frac{1}{2}
\theta''(\tau_{0})(\tau-\tau_{0})^{2}\right]\notag\\
&= r\mathcal{A}(\sigma)\sqrt{\frac{2\pi}{\theta''(\tau_{0})}}\,(D\tau_{0})^{-\frac{1}
{\sigma+2}}\exp[-\theta(\tau_{0})],
\label{eq_40}
\end{align}
provided $\theta''(\tau_{0})$ is large enough and where the double prime symbolizes a second
derivative of the $\theta(\tau)$ with respect to time $\tau$. Note that both conditions
refer to the intermediate asymptotic behavior in $x$. In the last step we complement
Eq.~(\ref{eq_40}) by substituting to it the coefficient $\mathcal{A}(\sigma)$ given by
Eq.~(\ref{eq_7}) and the function $\theta(\tau)$ together with its second derivative over
time at the minimum $\tau_{0}$ (see Eq.~(\ref{eq_39})). In this way we conclude that the
steady-state PDF for non-linear diffusion with the exponential resetting is
\begin{align}
p_{r}(x\!\mid\!x_{0})&\simeq A(\sigma)\left(\frac{r}{D}\right)^{\frac{2}{\sigma+4}}
\lvert x-x_{0}\rvert^{\frac{\sigma}{\sigma+4}}\,\mathrm{e}^{-\phi(x)},\;\;\text{with}
\label{eq_41}\\
\phi(x)&=B(\sigma)\left[\frac{r}{D}(x-x_{0})^{\sigma+2}\right]^{\frac{2}{\sigma+4}}
\notag,
\end{align}
where now the actual $\sigma$-dependent coefficients are
\begin{align}
A(\sigma)&=\frac{(2\pi)^{\frac{\sigma}{\sigma+4}}}{\sqrt{\sigma+4}}\left[\frac{
\Gamma(\Resize{1.8mm}{\frac{1}{\sigma}}+\Resize{1.8mm}{1})}{\sqrt{\sigma}\,
\Gamma(\Resize{1.8mm}{\frac{1}{\sigma}}+\Resize{1.8mm}{\frac{3}{2}})}\right]^{
\frac{\sigma-4}{\sigma+4}},\,\,\text{and}\label{eq_42}\\
B(\sigma)&=\frac{1}{2}\left(\frac{\sigma+4}{\sigma+2}\right)\left[\sqrt{\frac{2\pi}{
\sigma}}\frac{\Gamma(\Resize{1.8mm}{\frac{1}{\sigma}}+\Resize{1.8mm}{1})}
{\Gamma(\Resize{1.8mm}{\frac{1}{\sigma}}+\Resize{1.8mm}{\frac{3}{2}})}\right]^{
\frac{2\sigma}{\sigma+4}}\label{eq_43}.
\end{align}
\begin{figure}[t]
\centering
\includegraphics[scale=0.3]{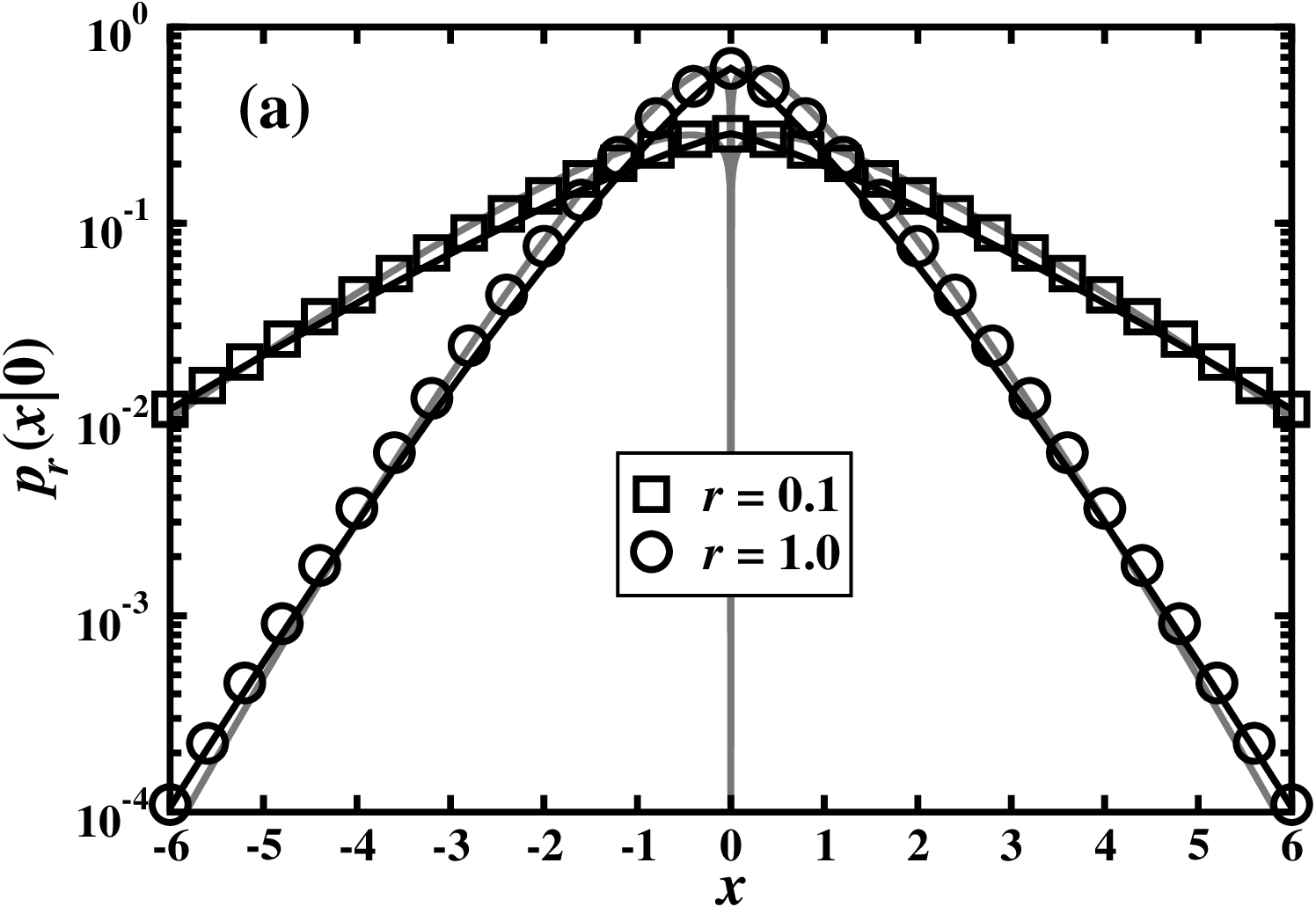}
\hspace{0.5cm}
\includegraphics[scale=0.3]{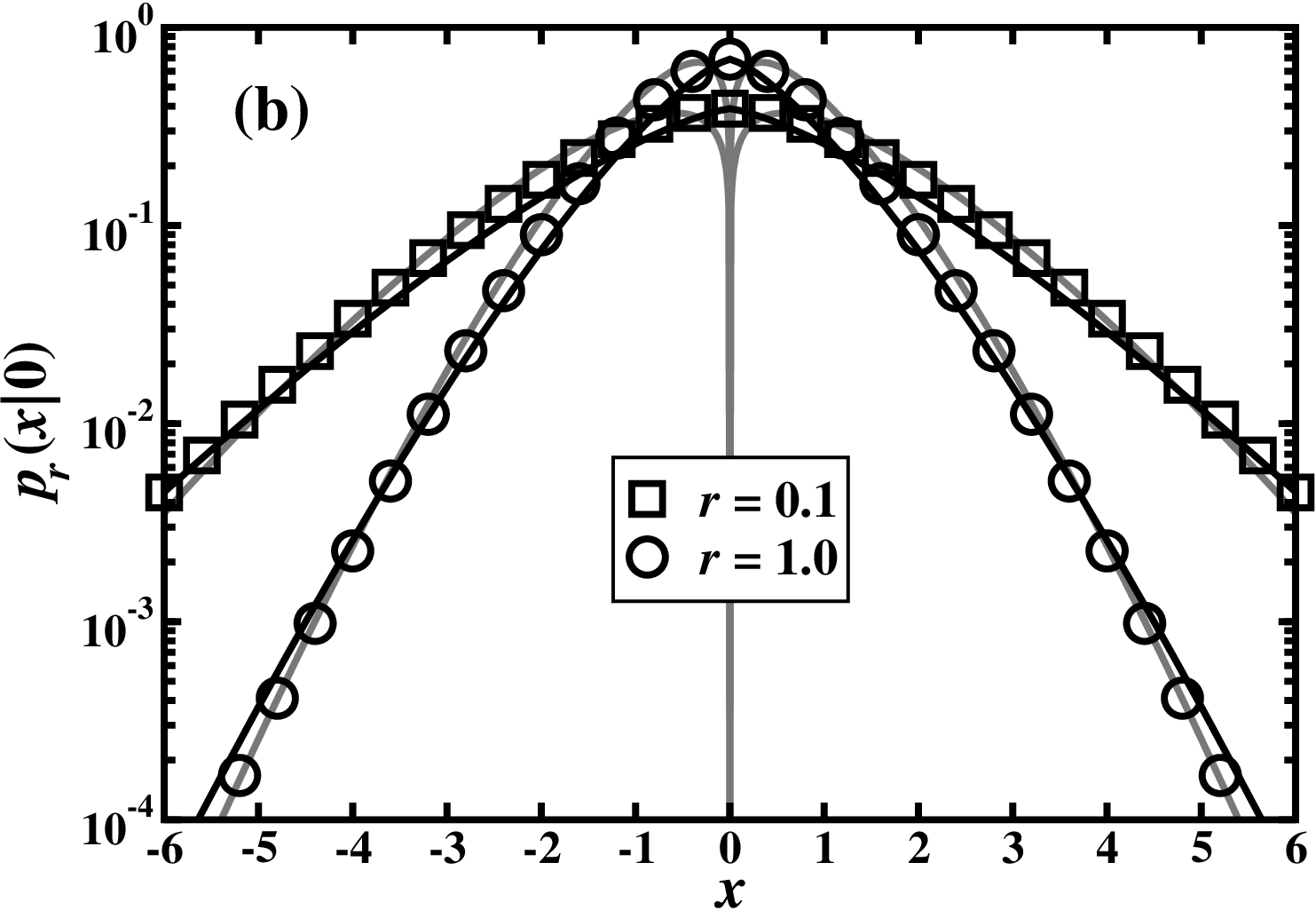}
\caption{The steady-state PDFs versus the position for two values of the parameter
$\sigma\!=\!1.0$ (a) and $2.0$ (b) and two chosen resetting rates $r\!=\!0.1$ (squares) and
$1.0$ (circls). Results of numerical integration conducted in Eq.~(\ref{eq_38}) are shown by
points. The solid black lines correspond to the PDFs described by Eq.~(\ref{eq_41}) with the
pre-exponential factor replaced by the expression obtained in Eq.~(\ref{eq_44}), whereas the
gray lines represent the PDF given by Eq.~(\ref{eq_44}). The fixed values of the diffusion
coefficient $D\!=\!1.0$ and the initial position $x_{0}\!=\!0$ of a diffusing particle have
been assumed.}
\label{fig5}
\end{figure}

As $\sigma\rightarrow0$ these two coefficients take the constant values, respectively,
$A(0)\!=\!\frac{1}{2}$ and $B(0)\!=\!1$. In a consequence we observe that unlike the formula
given by Eq.~(\ref{eq_35}) the stationary PDF in Eq.~(\ref{eq_41}) bowls down to the canonical
two-sided exponential probability distribution shown in Eq.~(\ref{eq_30}).

We underline once again that Eq.~(\ref{eq_41}) corresponds to the intermediate properties of
the PDF in $x$, so it ignores the case when $x\!=\!x_{0}$. In order to determine the actual
value of $p_{r}(x\!\mid\!x_{0})$ for $x\!=\!x_{0}$ we have to perform the integral in
Eq.~(\ref{eq_38}) with $\theta(\tau)\!=\!r\tau$ and $t\!\to\!\infty$. The result is
\begin{equation}
p_{r}(x_{0})=\mathcal{A}(\sigma)\left(\frac{r}{D}\right)^{\frac{1}{\sigma+2}}
\Gamma\left(\frac{\sigma+1}{\sigma+2}\right).
\label{eq_44}
\end{equation}

Figure~\ref{fig5} exemplifies the steady-state PDFs for the non-linear diffusion with the
Poissonian resetting. Here, we supposed two different values of the parameter $\sigma\!=\!1.0$
and $2.0$ and two chosen resetting rates $r\!=\!0.1$ and $1.0$ in the both cases. Due to these
plots we show that the data (points) obtained from the numerical integration performed on the
basis of Eqs.~(\ref{eq_37}) and (\ref{eq_38}) are in good agreement with the analytical results
(solid black lines) defined by Eq.~(\ref{eq_41}) in which the pre-exponential component was
replaced by the expression in Eq.~(\ref{eq_44}). The additional gray lines represent the
steady-state PDFs given by a whole function in Eq.~(\ref{eq_41}).

\section{Mean hitting time for non-linear diffusion with stochastic resetting}

\subsection{General method}

From now on we turn to the problem of finding the mean time needed for a diffusing particle
to reach a target localized at the origin $x\!=\!0$ of the semi-infinite interval if it starts
at a position $x\!=\!x_{0}$. The target is thought of as an immobile absorbing point so that
when the diffusing particle hits the origin for the first time it remains there forever. In
this sense our task is formally equivalent to the first-passage time problem or more precisely
the mean time to absorption (MTA). However, we additionally assume that during a diffusive
motion the particle is also reset to the initial position $x_{0}$ at a constant rate $r$ with
the probability $r\mathrm{e}^{-rt}$. The key quantity associated with the MTA is the survival
probability. As for the PDF in Eq.~(\ref{eq_21}) it is also possible to relate the survival
probability with the exponential resetting, $Q_{r}$, to that without resetting, $Q$. The
resulting equation reads
\begin{equation}
Q_{r}(t\!\mid\!x_{0})=\mathrm{e}^{-rt}Q(t\!\mid\!x_{0})+r\int_{0}^{t}\mathrm{d}
\tau\mathrm{e}^{-rt}Q(\tau\!\mid\!x_{0})Q_{r}(t-\tau\!\mid\!x_{0}),
\label{eq_45}
\end{equation}
where the first term on the right-hand site corresponds to trajectories without resets
whereas the second one represents trajectories in which resetting has occurred
\cite{Evans2020}. Note that the integral is performed with respect to the time $\tau$ that
elapsed since the last reset and there is a convolution of two survival probabilities. The
first one refers to the diffusive motion starting from the location $x_{0}$ in the absence
of resetting for duration $\tau$ and the second one concerns its continuation starting from
the same position with resetting up to time $t\!-\!\tau$. Using the Laplace transformation
for the survival probability
$\tilde{Q}_{r}(s\!\mid\!x_{0})\!=\!\int_{0}^{\infty}Q_{r}(t\!\mid\!x_{0})\mathrm{e}^{-st}
\mathrm{d}t$, we can convert Eq.~(\ref{eq_45}) to the appropriate equation in the Laplace
domain:
\begin{equation}
\tilde{Q}_{r}(s\!\mid\!x_{0})=\frac{\tilde{Q}(s+r\!\mid\!x_{0})}{1-r\,\tilde{Q}
(s+r\!\mid\!x_{0})}.
\label{eq_46}
\end{equation}
The algebraic form of the above equation is especially convenient to determine the MTA for
any diffusion process with the Poissonian resetting. Indeed, due to the relation between
MTA and the survival probability is defined in the following form,
$\overline{\tau}_{r}(x_{0}\!\rightarrow\!0)\!=\!\int_{0}^{\infty}Q_{r}(t\!\mid\!x_{0})
\mathrm{d}t$, we readily find that
\begin{equation}
\overline{\tau}_{r}(x_{0}\!\rightarrow\!0)=\lim_{s\to0}\int_{0}^{\infty}Q_{r}(t\!\mid\!x_{0})
\mathrm{e}^{-st}\mathrm{d}t=\lim_{s\to0}\tilde{Q}_{r}(s\!\mid\!x_{0}).
\label{eq_47}
\end{equation}
Therefore, connecting Eq.~(\ref{eq_46}) with this equation, we obtain
\begin{equation}
\overline{\tau}_{r}(x_{0}\!\rightarrow\!0)=\frac{\tilde{Q}(r\!\mid\!x_{0})}{1-r\,\tilde{Q}
(r\!\mid\!x_{0})}.
\label{eq_48}
\end{equation}
Note that determination of the MTA essentially comes down to the calculation of the survival
probability without exponential resetting in the Laplace domain where we finally need to
substitute $s\!=\!r$. For the linear diffusion taking place on the semi-infinite interval,
$0\leqslant x<\infty$, with the absorbing barrier at the origin $x\!=\!0$ and the initial
position at $x\!=\!x_{0}$, the survival probability
$Q(t\!\mid\!x_{0})\!=\!\int_{0}^{\infty}p(x,t\!\mid\!x_{0})\mathrm{d}x$. In this formula the
integral is performed with respect to the PDF being the solution of a linear partial
differential equation with the absorbing boundary condition $p(0,t\!\mid\!x_{0})\!=\!0$.
A conventional technique allowing to solve this problem is the method of images which emerges
from a more general method of Green's functions. They ensure the PDF must vanish at the
absorbing point at any time. However, the both methods and especially this type of an
absorbing boundary condition are improper regarding the non-linear diffusion equation and
can not be applied in this case. We intend to deal with this problem in a separate paper.
While now, we will employ the alternative approach based on a system of two related equations.

The first equation determines a relation between the survival probability $Q(t\!\mid\!x_{0})$
and the first-passage time distribution $F(t\!\mid\!x_{0})$ expressed in the form of the
ordinary differential equation
\begin{equation}
\frac{\mathrm{d}Q(t\!\mid\!x_{0})}{\mathrm{d}t}=-F(t\!\mid\!x_{0}),
\label{eq_49}
\end{equation}
where the survival probability satisfies the initial boundary condition,
$Q(0\!\mid\!x_{0})\!=\!1$. It means that a diffusing particle definitely exists (survives) at
$t\!=\!0$. The second equation has a form of the integral equation in which the first-passage
time distribution is related to the PDF as follows
\begin{equation}
p(0,t\!\mid\!x_{0})=\int_{0}^{t}F(\tau\!\mid\!x_{0})\,p(0,t-\tau\!\mid\!0)\,\mathrm{d}\tau.
\label{eq_50}
\end{equation}
This equation defines the PDF or the propagator from $x_{0}$ to $0$ for a random dynamics
as an integral over the first time to reach the point $0$ at a time $\tau\!\leqslant\!t$
followed by a loop from $(0,\tau)$ to $(0,t)$ in the remaining time $t\!-\!\tau$. Upon
integrating Eq.~(\ref{eq_49}) in the time range from $0$ to $t$ and accounting for the
initial condition for the survival probability we obtain
\begin{equation}
Q(t\!\mid\!x_{0})=1-\int_{0}^{t}F(\tau\!\mid\!x_{0})\,\mathrm{d}\tau.
\label{eq_51}
\end{equation}
In terms of the Laplace transforms of the survival and first-passage time probabilities
the above formula becomes the algebraic equation
\begin{equation}
\tilde{Q}(s\!\mid\!x_{0})=\frac{1}{s}\left[1-\tilde{F}(s\!\mid\!x_{0})\right].
\label{eq_52}
\end{equation}
Because the integral on the right-hand side of Eq.~(\ref{eq_50}) is a convolution so we have
to use again the Laplace transform in order to convert this equation into the algebraic form:
\begin{equation}
\tilde{F}(s\!\mid\!x_{0})=\frac{\tilde{p}(0,s\!\mid\!x_{0})}{\tilde{p}(0,s\!\mid\!0)}.
\label{eq_53}
\end{equation}
Combining the last two expressions yields the relation between the survival probability and
the PDFs in the Laplace domain:
\begin{equation}
\tilde{Q}(s\!\mid\!x_{0})=\frac{1}{s}\left[1-\frac{\tilde{p}(0,s\!\mid\!x_{0})}
{\tilde{p}(0,s\!\mid\!0)}\right],
\label{eq_54}
\end{equation}
whereas inserting the above formula into Eq.~(\ref{eq_48}) results in the MTA for a random
process undergoing the exponential (Poissonian) resetting:
\begin{equation}
\overline{\tau}_{r}(x_{0}\!\rightarrow\!0)=\frac{1}{r}\left[\frac{\tilde{p}(0,r\!\mid\!0)}
{\tilde{p}(0,r\!\mid\!x_{0})}-1\right].
\label{eq_55}
\end{equation}
Let us emphasize once more that the survival probability recovered form Eq.~(\ref{eq_54}) by
the inverse Laplace transform needs to satisfy the appropriate conditions, that is
$Q(0\!\mid\!x_{0})\!=\!1$ and additionally $Q(\infty\!\mid\!x_{0})\!=\!0$ in the limit of
infinitely long times.

As a simple example we first examine Eq.~(\ref{eq_55}) in the context of the linear
diffusion for which the PDF or more generally the propagator from an initial position
in $x'$ to the final location in $x$ is given by the Gaussian distribution
\begin{equation}
p(x,t\!\mid\!x')=\frac{1}{\sqrt{4\pi Dt}}\exp\left[-\frac{(x-x')^{2}}{4Dt}\right].
\label{eq_56}
\end{equation}
Taking its Laplace transform yields
\begin{equation}
\tilde{p}(x,s\!\mid\!x')=\frac{1}{2\sqrt{D s}}
\exp\left(-\sqrt{\frac{s}{D}}\mid\!x-x'\!\mid\right).
\label{eq_57}
\end{equation}
A direct substitution of this expression into Eq.~(\ref{eq_55}) along with $s\!=\!r$ leads
to the seminal result for the MTA from $x_{0}$ to $0$, namely
\begin{equation}
\overline{\tau}_{r}(x_{0}\!\rightarrow\!0)=\frac{1}{r}\left[\exp\left(\sqrt{\frac{r}{D}}\,
x_{0}\right)-1\right].
\label{eq_58}
\end{equation}
This formula was derived for the first time in the reference \cite{Evans2011}. However, the
method used in that paper was completely different because consisting in the solution of
a backward master equation for the survival probability with appropriate boundary and
initial conditions. The result exposed in Eq.~(\ref{eq_58} can also be received by applying
the method of images in which the boundary condition $p(0,t\!\mid\!x_{0})\!=\!0$
corresponding to the absorbing barrier in the position $x\!=\!0$ is of a principal
meaning. But, as we have mentioned earlier, this method is inappropriate and hence useless
in application to the non-linear diffusion equation. Instead, we will employ in our further
studies the alternative approach based on Eqs.~(\ref{eq_54}) and (\ref{eq_55}) which applies
to any diffusion process.

\subsection{Results for non-linear diffusion}

The key quantity in Eqs.~(\ref{eq_54}) and (\ref{eq_55}) is the Laplace transform of the
PDF, or more precisely, the propagator $\tilde{p}(0,s\!\mid\!x')$ for a free non-linear
diffusion with $x'$ being the initial $x_{0}$ or final position $0$. In formal terms, the
problem of finding this quantity comes down to the calculation of a definite integral in
Eq.~(\ref{eq_28}) (see also Eq.~(\ref{eq_27}) resulting in the PDF for the non-linear
diffusion affected by the exponential resetting. Thus the Laplace transform of the
propagator is as follows:
\begin{equation}
\tilde{p}(0,s\!\mid\!x')=\int_{\eta(x',\sigma)}^{\infty}\frac{\mathcal{A}(\sigma)}
{(D t)^{\frac{1}{\sigma+2}}}\left[1-\left(\frac{\eta(x',\sigma)}{D t}\right)^{\frac{2}
{\sigma+2}}\right]^{\frac{1}{\sigma}}\exp(-st)\,\mathrm{d}t,
\label{eq_59}
\end{equation}
where the lower limit of integration is determined by the auxiliary function
$\eta(x',\sigma)\!=\!(\sqrt{\mathcal{B}(\sigma)}\mid\!x'\!\mid)^{\sigma+2}$. Notice a
lack of the resetting rate constant $r$ in front of the integral compared to
Eq.~(\ref{eq_28}). We already know from Sec.~3.2 that performing the precise integration in
Eq.~(\ref{eq_59}) for any value of the parameter $\sigma\!>\!0$ is a great challenge.
Nevertheless, there exists two exceptional cases when this operation is possible. In the
simpler case for $\sigma\!=\!1$, the Laplace transform of the propagator in Eq.~(\ref{eq_59}
is
\begin{equation}
\tilde{p}(0,s\!\mid\!x')=\left(\frac{3}{32}\right)^{\frac{1}{3}}(\sqrt{D}s)^{-\frac{2}{3}}
\left[\Gamma\left(\frac{2}{3},\frac{2\!\mid\!x'\!\mid^{3}}{9D}s\right)+
\left(\frac{2s}{9D}\right)^{\frac{2}{3}}x'^{2}\,\mathrm{Ei}\left(\frac{2\!\mid\!x'\!\mid^{3}}
{9D}s\right)\right],
\label{eq_60}
\end{equation}
where, as in Eq.~(\ref{eq_31}), $\Gamma(\alpha,z)$ and $\mathrm{Ei}(z)$ denote, respectively,
the upper incomplete gamma function and the exponential integral function. The second exact
result is possible to be obtained for the parameter $\sigma\!=\!2$. Using the procedure
described in Appendix we readily get from Eq.~(\ref{eq_59}) that the propagator
\begin{equation}
\tilde{p}(0,s\!\mid\!x')=\frac{\pi\!\mid\!x'\!\mid^{3}}{8\sqrt{2}D}\exp\!\left(-
\frac{\pi^{2}x'^{4}}{32D}s\right)\left[\mathrm{K}_{\frac{3}{4}}\!\left(\frac{
\pi^{2}x'^{4}}{32D}s\right)-\mathrm{K}_{\frac{1}{4}}\!\left(\frac{\pi^{2}x'^{4}}{32D}s
\right)\right],
\label{eq_61}
\end{equation}
where $\mathrm{K}_{\nu}(z)$ is the modified Bessel function of the second kind.

At this point an obvious question arises about the remaining values of the parameter $\sigma$
for which the integral in Eq.~(\ref{eq_59}) is tractable at least in approximation. In
Sec.~3.2 we have proposed such a method consisting in the use of two solutions based on
the exponential and algebraic approximations. By contrast with the first approximation, the
algebraic approach allows us to derive the PDF in Eq.~(\ref{eq_35}) manifesting much
better consistency with the numerical and two exact results involved in Eqs.~(\ref{eq_31})
and (\ref{eq_32}). For this reason, we will consequently employ this approximation
throughout the rest of our analysis. Therefore by conducting similar calculations as
in Sec.~3.2, we infer from Eq.~(\ref{eq_59}) that the Laplace transform of the propagator
\begin{equation}
\tilde{p}(0,s\!\mid\!x')\simeq\frac{\mathcal{A}(\sigma)}{D^{\frac{1}{\sigma+2}}
s^{\frac{\sigma+1}{\sigma+2}}}\Gamma\left(\frac{\sigma+1}{\sigma+2},\frac{s}{D}
\eta(x',\sigma)\right)\left[1-\left(\frac{s}{D}\eta(x',\sigma)\right)^{\frac{2}
{\sigma+2}}\frac{\Gamma\left(\frac{\sigma-1}{\sigma+2},\frac{s}{D}\eta(x',\sigma)
\right)}{\Gamma\left(\frac{\sigma+1}{\sigma+2},\frac{s}{D}\eta(x',\sigma)\right)}
\right]^{\frac{1}{\sigma}}.
\label{eq_62}
\end{equation}
In the following we will show that this expression when inserted in Eq.~(\ref{eq_55}) leads
to the approximate formula for the MTA which turns out to be an unexpectedly accurate result
provided that the parameter $\sigma$ falls in the range between $1.0$ and $2$. This means
that the analytical formula for the MTA is in good agreement with the data received due to
the numerical integration carried out in Eq.~(\ref{eq_59}) and used as the input into
Eq.~(\ref{eq_55}). However, an intriguing question still remains open whether it is possible
to construct an appropriate expression for MTA if the positive parameter $\sigma\!<\!1$ and
$\sigma\!>2$. We consciously postpone this problem to the future research.

Before we proceed to a determination of MTA, let us first examine whether the survival
probability resulting form the combination of Eqs.~(\ref{eq_54}) with (\ref{eq_62}) for the
non-linear diffusion satisfies the same boundary conditions as it is the case with the
ordinary diffusive motion in a presence of the absorbing barrier. Note if $x'\!=\!0$ then
the auxiliary function in Eq.~(\ref{eq_62}) disappears, namely $\eta(0,\sigma)\!=\!0$, and
hence each incomplete gamma function in this equation becomes the Euler gamma function.
Taking into account this property in Eq.~(\ref{eq_54}) we obtain the appropriate formula
for the Laplace transform of the survival probability in the case of the free non-linear
diffusion:
\begin{equation}
\tilde{Q}(s\!\mid\!x_{0})\simeq\frac{1}{s}\left\{1-
\frac{\Gamma\left(\frac{\sigma+1}{\sigma+2},\frac{s}{D}\eta(x_{0},\sigma)\right)}{
\Gamma\left(\frac{\sigma+1}{\sigma+2}\right)}
\left[1-\left(\frac{s}{D}\eta(x_{0},\sigma)\right)^{\frac{2}
{\sigma+2}}\frac{\Gamma\left(\frac{\sigma-1}{\sigma+2},\frac{s}{D}\eta(x_{0},\sigma)
\right)}{\Gamma\left(\frac{\sigma+1}{\sigma+2},\frac{s}{D}\eta(x_{0},\sigma)\right)}
\right]^{\frac{1}{\sigma}}\right\},
\label{eq_63}
\end{equation}
where $\eta(x_{0},\sigma)\!=\!(\sqrt{\mathcal{B}(\sigma)}\mid\!x_{0}\!\mid)^{\sigma+2}$
depends solely on the initial position $x_{0}$ of a particle and the parameter $\sigma$.
Due to the fact that we have not for our disposal the inverse Laplace transform for the
above survival probability, we employ the adequate limit theorems to validate the boundary
conditions for this quantity. The first proposition for the initial condition is that if
$t\!=\!0$ then
$Q(0\!\mid\!x_{0})\!=\!\lim_{t\to0}Q(t\!\mid\!x_{0})\!=\!\lim_{s\to\infty}s\,\tilde{Q}
(s\!\mid\!x_{0})$, whereas the second condition corresponds to the limit when
$t\!\to\!\infty$ and states that
$Q(\infty\!\mid\!x_{0})\!=\!\lim_{t\to\infty}Q(t\!\mid\!x_{0})\!=\!
\lim_{s\to0}s\,\tilde{Q}(s\!\mid\!x_{0})$ \cite{Schiff1999}. In addition, it is suffice
to note that $\lim_{z\to0}\Gamma(\alpha,z)\!=\!\Gamma(\alpha)$ and the asymptotic expansion
of the upper incomplete gamma function $\Gamma(\alpha,z)\!\propto\!z^{\alpha-1}
\mathrm{e}^{-z}$ for $\mid\!z\!\mid\!\to\!\infty$. Then, by virtue of the limit theorems
applied to Eq.~(\ref{eq_61}) we immediately conclude that $Q(0\!\mid\!x_{0})\!=\!1$ and
$Q(\infty\!\mid\!x_{0})\!=\!0$. It is not an especially difficult task to verify a
correctness of these boundary conditions for survival probabilities derived from
Eq.~(\ref{eq_54}) along with Eqs.~(\ref{eq_60}) and (\ref{eq_61}) where the parameter
$\sigma\!=\!1$ and $2$, respectively.

From Eq.~(\ref{eq_63}) we see that MTA for the free non-linear diffusion, that is
$\overline{\tau}(x_{0}\!\to\!0)\!=\!\lim_{s\to0}\tilde{Q}(s\!\mid\!x_{0})$, is evidently
divergent. This tendency is consistent with the ordinary diffusion on a semi-infinite
interval with the absorbing barrier at the origin \cite{Redner2001}. Such a behavior of
the MTA was also confirmed in a more general variant of the fractional heterogeneous
non-linear diffusion equation\cite{Wang2008}. Nevertheless, we want to strongly emphasize
that the method used in this paper may raise serious doubts because of an application of
the standard boundary condition according to which the PDF disappears in the absorbing
point. This problem requires a deeper analysis and we intend to consider it in a separate
paper.
\begin{figure}[t]
\centering
\includegraphics[scale=0.35]{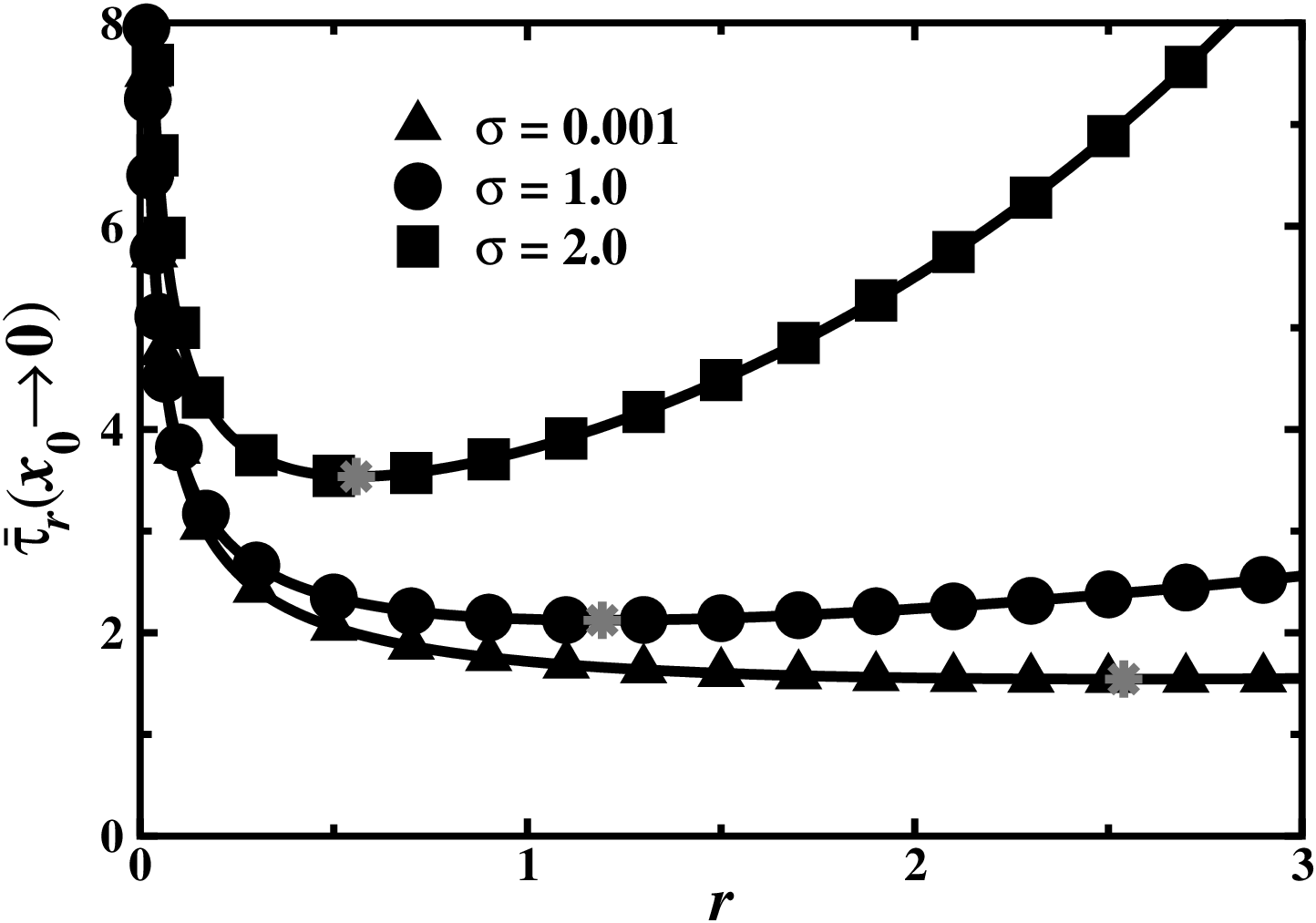}
\caption{The MTA $\overline{\tau}_{r}$ (the mean first-passage time) versus the resetting
rate $r$ for different values of the parameter $\sigma$. The points marked by circles,
squares and triangles are outcomes obtained in two stages of computations consisting of
the numerical integration performed in Eq.~(\ref{eq_59}) followed by the insertion of the
resulting data into Eq.~(\ref{eq_55}). The solid lines represent the exact analytical
results given by Eqs.~(\ref{eq_58}), (\ref{eq_64}) and (\ref{eq_65}). The grey stars
indicate the locci where MTA attains the minimum value. The fixed values of the diffusion
coefficient $D\!=\!1.0$ and the initial position $x_{0}\!=\!1.0$ of a diffusing particle
have been assumed.}
\label{fig6}
\end{figure}
\begin{figure}[t]
\centering
\includegraphics[scale=0.35]{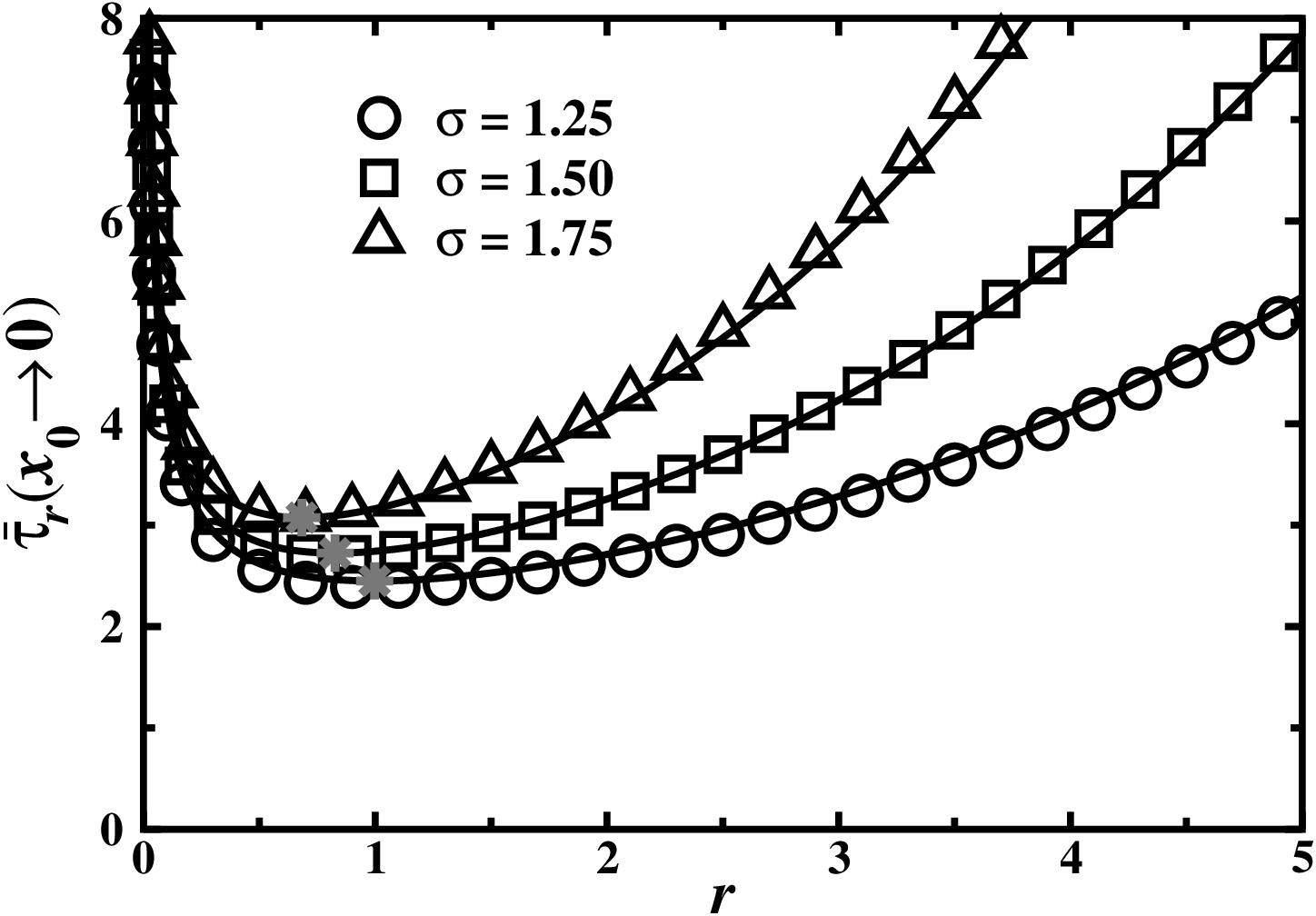}
\caption{The MTA $\overline{\tau}_{r}$ (the mean first-passage time) versus the resetting
rate $r$ for different values of the parameter $\sigma$ falling in the range from $1.0$ to
$2.0$. The points marked by circles, squares and triangles are outcomes obtained in two
stages of computations consisting of the numerical integration performed in Eq.~(\ref{eq_59})
followed by the insertion of the resulting data into Eq.~(\ref{eq_55}). The solid lines
represent the approximate analytical result shown in Eq.~(\ref{eq_66}). The grey stars
indicate the locci where MTA attains the minimum value. The fixed values of the diffusion
coefficient $D\!=\!1.0$ and the initial position $x_{0}\!=\!1.0$ of a diffusing particle
have been assumed.}
\label{fig7}
\end{figure}
A propensity of the MTA to be infinite substantially changes when the non-linear diffusion
occurs under the influence of exponential resets. To show this effect we can proceed in two
equivalent ways. According to the first way it is enough to substitute the Laplace transform
of the survival probability (see Eq.~(\ref{eq_54}) with $s\!=\!r$ obtained, respectively,
form Eq.~(\ref{eq_60}) for $\sigma\!=\!1$, Eq.~(\ref{eq_61}) for $\sigma\!=\!2$ and
(\ref{eq_62}) for $1\!<\!\sigma\!<\!2$ (see also Eq.~(\ref{eq_63})) into Eq.~(\ref{eq_48}).
In the second case we simply insert the Laplace transforms of PDF given by Eqs.~(\ref{eq_60}),
(\ref{eq_61}) and (\ref{eq_62}) into the formula derived in Eq.~(\ref{eq_55}). The final
results read
\begin{equation}
\overline{\tau}_{r}(x_{0}\!\rightarrow\!0)=\frac{1}{r}\left[\frac{
\Gamma\left(\frac{2}{3}\right)}{\Gamma\left(\frac{2}{3},\frac{2\mid x_{0}\mid^{3}}{9D}r
\right)+\left(\frac{2r}{9D}\right)^{\frac{2}{3}}x_{0}^{2}\,\mathrm{Ei}\left(
-\frac{2\mid x_{0}\mid^{3}}{9D}r\right)}-1\right],
\label{eq_64}
\end{equation}
for the parameter $\sigma\!=\!1$,
\begin{equation}
\overline{\tau}_{r}(x_{0}\!\rightarrow\!0)=\frac{1}{r}\left\{\frac{
16\left(\frac{D}{r}\right)^{\frac{3}{4}}\exp\left(\frac{\pi^{2}x_{0}}{32D}r\right)}{
\sqrt{\pi}\mid\!x_{0}\!\mid^{3}\Gamma\left(\frac{1}{4}\right)\left[\mathrm{K}_{\frac{3}{4}}
\left(\frac{\pi^{2}x_{0}}{32D}r\right)-\mathrm{K}_{\frac{3}{4}}
\left(\frac{\pi^{2}x_{0}}{32D}r\right)\right]}-1\right\},
\label{eq_65}
\end{equation}
for the parameter $\sigma\!=\!2$, and
\begin{equation}
\overline{\tau}_{r}(x_{0}\!\rightarrow\!0)\simeq\frac{1}{r}\left\{
\frac{\Gamma\left(\frac{\sigma+1}{\sigma+2}\right)}{
\Gamma\left(\frac{\sigma+1}{\sigma+2},\frac{r}{D}\eta(x_{0},\sigma)\right)}
\left[1-\left(\frac{r}{D}\eta(x_{0},\sigma)\right)^{\frac{2}
{\sigma+2}}\frac{\Gamma\left(\frac{\sigma-1}{\sigma+2},\frac{r}{D}\eta(x_{0},\sigma)
\right)}{\Gamma\left(\frac{\sigma+1}{\sigma+2},\frac{r}{D}\eta(x_{0},\sigma)\right)}
\right]^{-\frac{1}{\sigma}}-1\right\},
\label{eq_66}
\end{equation}
for the parameter $\sigma$ in the range between $1.0$ and $2.0$, where the auxiliary function
$\eta(x_{0},\sigma)\!=\!(\sqrt{\mathcal{B}(\sigma)}\mid\!x_{0}\!\mid)^{\sigma+2}$.
Dependencies of the MTA $\overline{\tau}_{r}(x_{0}\!\rightarrow\!0)$ on the resetting rate
$r$ described by Eqs.~(\ref{eq_58}), (\ref{eq_64}) and (\ref{eq_65}) are drawn in
Fig.~\ref{fig6} in the form of solid lines. For comparison all the points collected on this
plot symbolize the data obtained firstly form numerical integration conducted in
Eq.~(\ref{eq_59}) for $\sigma\!=\!0.001$, $1.0$ and $2.0$ and next entered as the
input data into Eq.~(\ref{eq_55}). The agreement of these results confirms the numerical
procedure we have used to compute MTA works without blemish. The same relations expressed by
the approximate Eq.~(\ref{eq_66}) for disparate values of the parameter $\sigma\!=\!1.25$,
$1.5$ and $1.75$ are shown in Fig.~\ref{fig7}. Note that in both graphs the MTA diverges as
$r\!\to\!0$ which retrieves the well-known result that the mean first-passage time to reach
the origin tends to infinity when a particle diffuses on the semi-infinite interval in the
absence of resetting. On the other hand, the MTA also diverges as $r\!\to\!\infty$, because
as the resetting rate increases the diffusing particle does not have enough time, i.e. it
has less time between resets, to reach the origin. Therefore, we infer form the behavior of
the MTA as a function of $r$ shown in Figs.~\ref{fig6} and \ref{fig7} that there exists
an optimal rate $r_{\ast}$ that minimizes this function. To determine its value we have to
solve the following equation, $\left(\frac{\mathrm{d}\overline{\tau}_{r}}{\mathrm{d}r}\right)
_{r\!=\!r_{\ast}}\!=\!0$, starting from determination of the first derivative of the MTA with
respect to the resetting rate. While this is not too difficult task for Eq.~(\ref{eq_58}),
it becomes more complex for the rest of formulae given by Eqs.~(\ref{eq_64}), (\ref{eq_65})
and (\ref{eq_66}). Therefore, instead conducting analytical calculations in order to find
the optimal resetting rate $r_{\ast}$ we have solved this problem numerically. The values of
$r_{\ast}$ for which MTA attains the minimum are distinguished in Figs.~\ref{fig6} and
\ref{fig7} by grey asterisks. Figure~\ref{fig8} presents how the optimal resetting rate
$r_{\ast}$ depends on the parameter $\sigma$ for various distances $x_{0}$ from the initial
position to the target localized at the origin of the semi-infinite interval (a) and the
very distance $x_{0}$ for selected values of the parameter $\sigma$ (b). In both cases the
monotonic decrease of $r_{\ast}$ with increase of $x_{0}$ and $\sigma$ is observed such as
for the scaled Brownian motion \cite{Sokolov2019}.
\begin{figure}[t]
\centering
\includegraphics[scale=0.31]{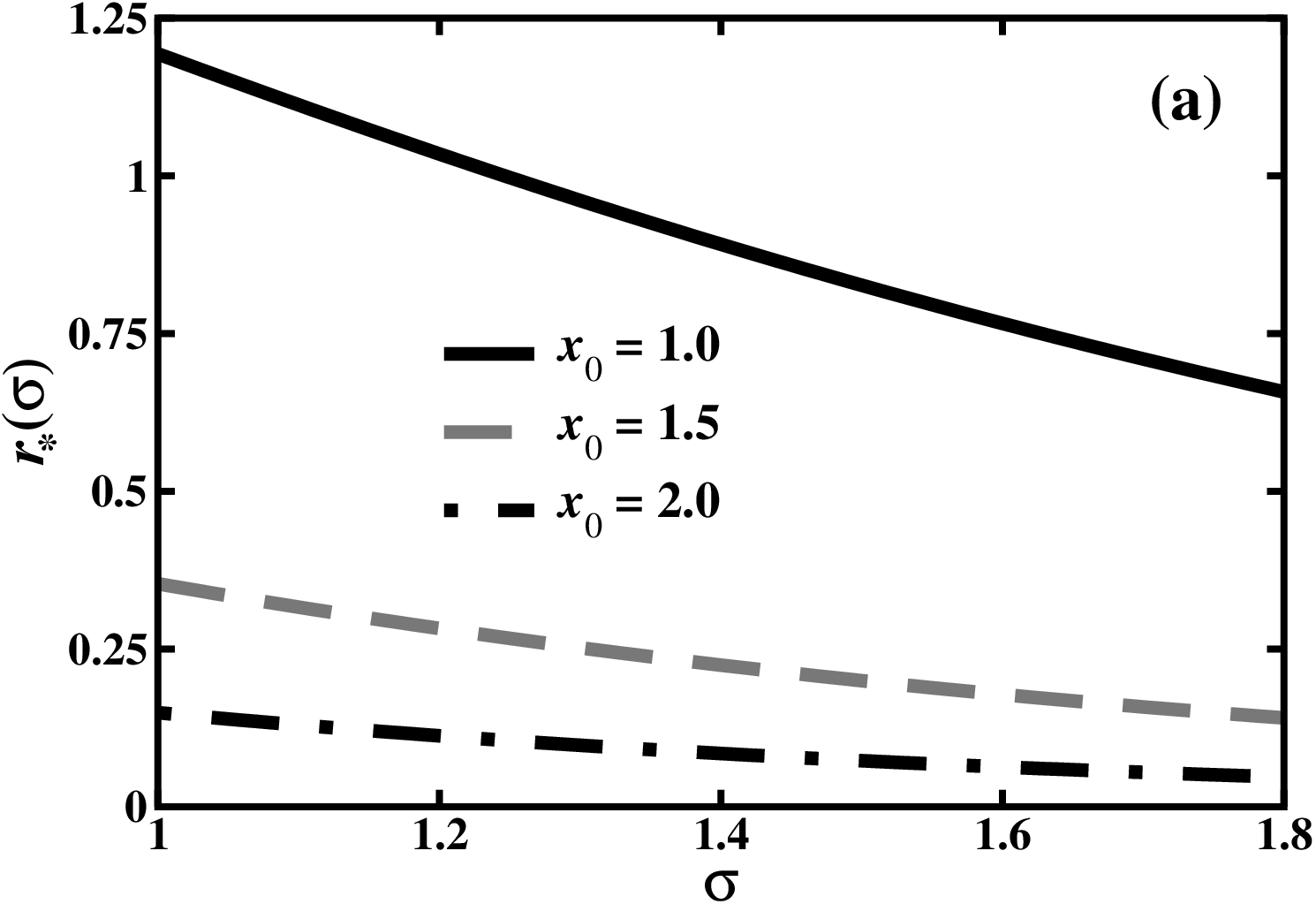}
\hspace{0.5cm}
\includegraphics[scale=0.31]{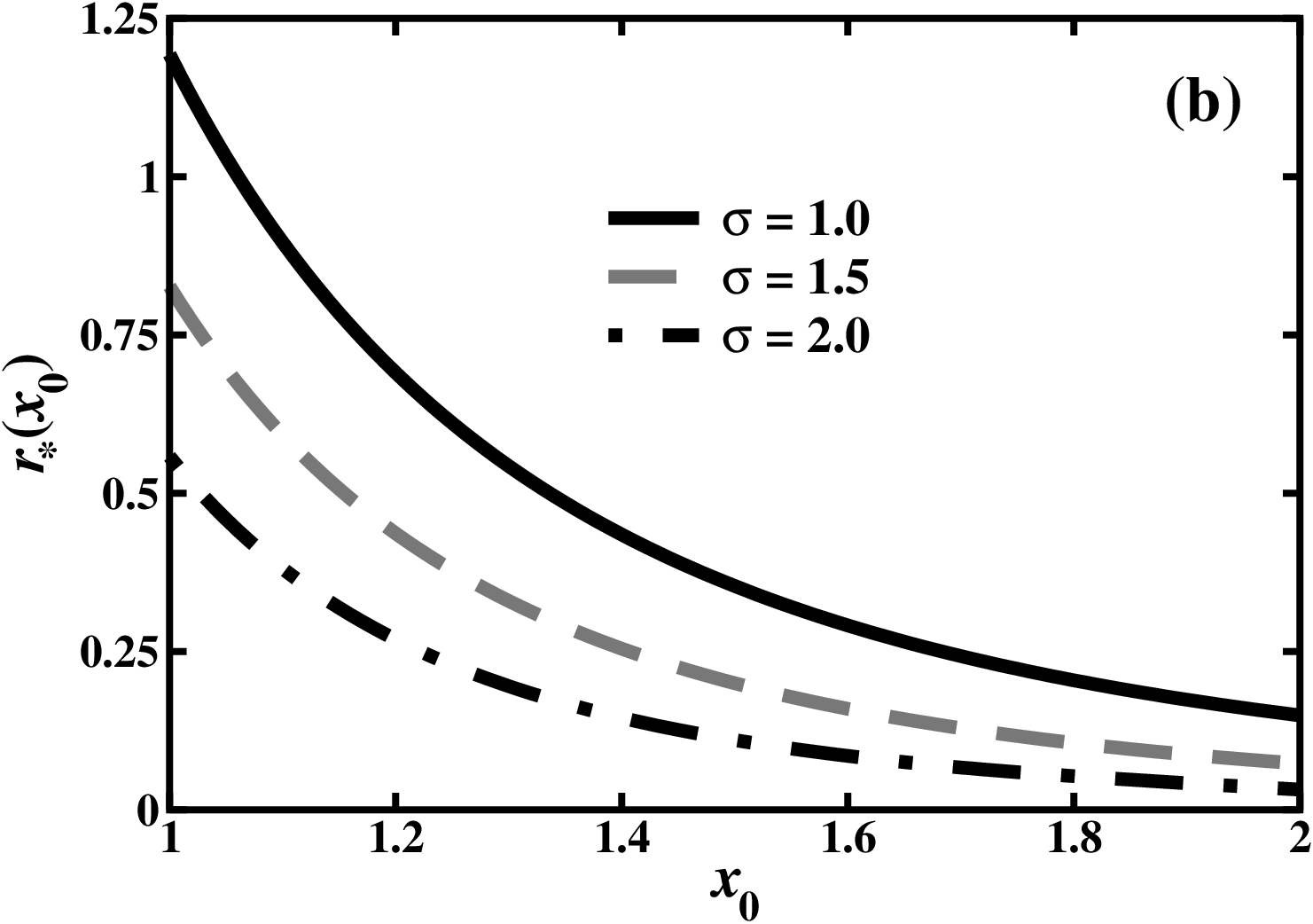}
\caption{The optimal resetting rate $r_{\ast}$ versus the parameter $\sigma$ (a) and
the distance $x_{0}$ to the target (absorbing point) located at the origin of a semi-infinite
interval. In both cases we have assumed the fixed value of the diffusion coefficient
$D\!=\!1.0$.}
\label{fig8}
\end{figure}
\begin{figure}[b]
\centering
\vspace{1.05cm}
\includegraphics[scale=0.35]{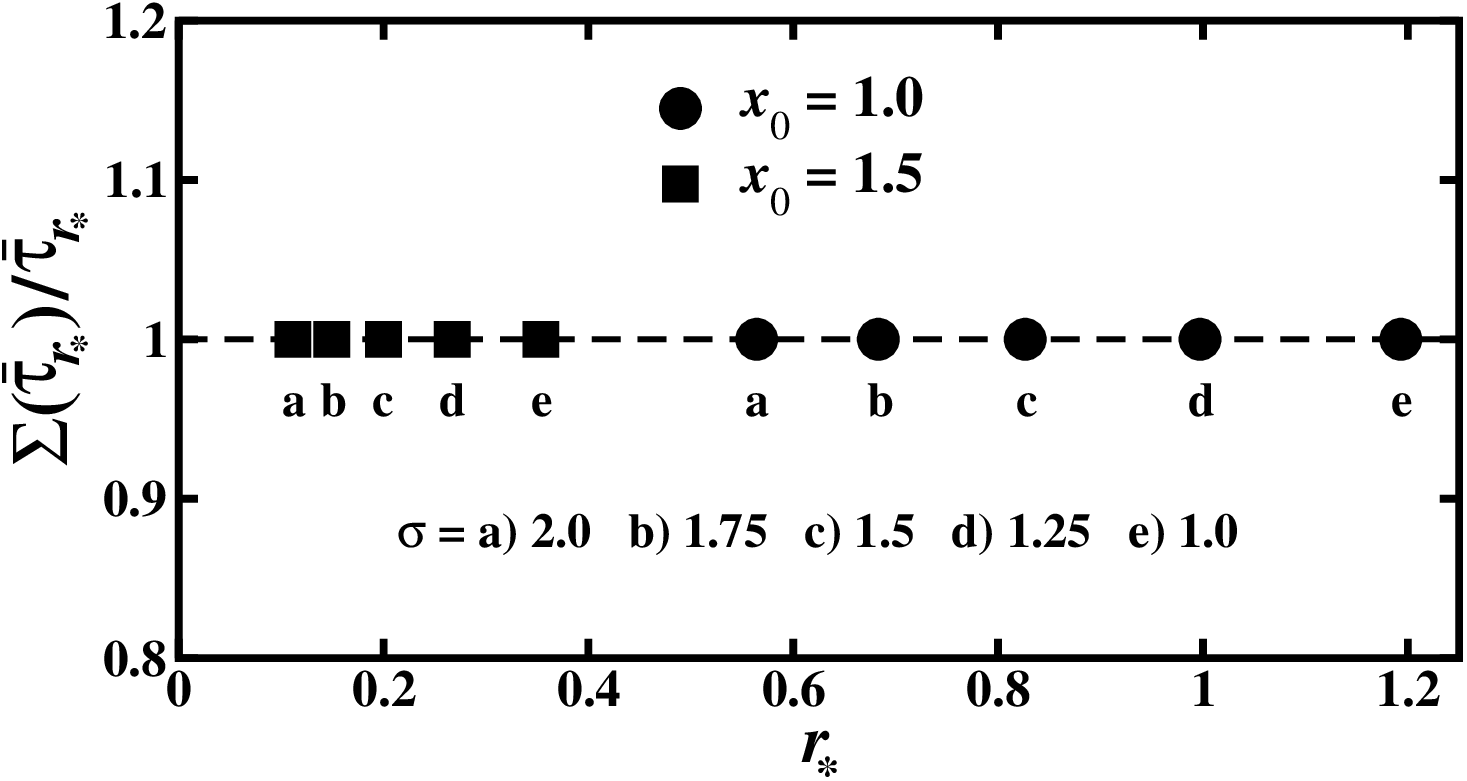}
\caption{Relative fluctuation in the MTA (the mean first-passage time) versus the optimal
resetting rate for different values of the parameter $\sigma$ and two chosen distances $1.0$
and $1.5$ from the initial position $x_{0}$ to the target located at the origin $x\!=\!0$ of
the semi-infinite interval. The collected numerical results confirm the assertion expressed
in Eq.~(\ref{eq_67}). Here, we have assumed the fixed value of the diffusion coefficient
$D\!=\!1.0$.}
\label{fig9}
\end{figure}

One of outstanding results concerning any random process intermittent by the exponential
resetting is that the relative fluctuation in the MTA, or the mean first-passage time, of an
optimally restarted process is always unity \cite{Reuveni2016b}. To be more specific, if
the resetting rate $r\!=\!r_{\ast}$, i.e. it is equal to the optimal resetting rate, then the
relative standard deviation in the MTA satisfies
\begin{equation}
\frac{\Sigma(\overline{\tau}_{r_{\ast}})}{\overline{\tau}_{r_{\ast}}}=1,
\label{eq_67}
\end{equation}
and this relation has been proven to be universal. Here, the quantity
$\Sigma(\overline{\tau}_{r})\!=\!\sqrt{\,\overline{\tau^{2}_{r}}-
\overline{\tau}_{r}^2}$ is the standard deviation of the mean first-passage time and
$\overline{\tau^{2}_{r}} \!=\!-2\left[\frac{\mathrm{d}\tilde{Q}_{r}(s\,\mid\,x_{0})}
{\mathrm{d}s}\right]_{s=0}$ defines its second moment. In this formula the Laplace
transform of the survival probability $\tilde{Q}_{r}(s\mid x_{0})$ is determined
in Eq.~(\ref{eq_46}). So, the natural question arises whether the relation in
Eq.~(\ref{eq_55}) is also satisfied for the non-linear diffusion with the exponential
resetting. To give answer to this specific question we proceed as follows. Let us recall
the Laplace transform of the survival probability for the non-linear diffusion in the
absence of resetting is contained in Eq.~(\ref{eq_61}). By virtue of the second moment of
the MTA including exponential restarts along with Eq.~(\ref{eq_46}) we get that
\begin{equation}
\overline{\tau^{2}_{r}}(x_{0}\to0)=-\frac{2\left[\frac{\mathrm{d}}{\mathrm{d}s}\tilde{Q}
(s\!\mid\!x_{0})\right]_{s=r}}{[1-r\tilde{Q}(r\!\mid\!x_{0})]^{2}}.
\label{eq_68}
\end{equation}
Combining the above formula with the MTA in Eq.~(\ref{eq_48}) and its squared counterpart
we easily calculate the relative standard deviation of the MTA, which is as follows:
\begin{equation}
\frac{\Sigma(\overline{\tau}_{r})}{\overline{\tau}_{r}}=\sqrt{-\bigg(\frac{2
\frac{\mathrm{d}}{\mathrm{d}s}\tilde{Q}(s\!\mid\!x_{0})\!\mid_{s=r}}{\tilde{Q}^{\,2}
(r\!\mid\!x_{0})}+1\bigg)}.
\label{eq_69}
\end{equation}
In the last step we could insert the survival probability defined by Eq.~(\ref{eq_43})
into this formula and check out validity of the proposition expressed in Eq.~(\ref{eq_66}).
Nevertheless, we will proceed in another way because the resulting formula is too complex
and the optimal resetting rates $r_{\ast}$ can only be determined by means of a numerical
procedure. Thus, we firstly determine their values for a few exemplary values of the
parameter $\sigma$ and different distances from the initial position $x_{0}$ to the origin
of the semi-infinite interval. Then, by performing numerical computations on the basis of
Eqs.~(\ref{eq_61}) and (\ref{eq_68}) we find the ratio
$\Sigma(\overline{\tau}_{r_{\ast}})/\overline{\tau}_{r_{\ast}}$. The plot shown in
Fig.~\ref{fig9} illustrates the final results and confirm correctness of the theorem in
Eq.~(\ref{eq_67}) for non-linear diffusion with the exponential resetting.

\section{Conclusions}

Despite evident differences between the free linear and non-linear diffusion, there exists one
common property connecting the both processes, that is, inability to reach the non-equilibrium
steady state. This situation dramatically changes when a particle executing a diffusive motion
undergoes the action of a permanent external force driving a system out of equilibrium. In this
paper we examined the alternative mechanism that enables a diffusing particle to be in the
stationary state. Instead of the force a system is subject to resetting, a kind of stochastic
process in which some moving object is turned back at random moments of time to a given initial
state, from where it starts its motion anew. Since the first publication considering this
problem \cite{Evans2011}, a lot of papers have appeared in recent years on different variants
of diffusion models with stochastic resetting. To our best knowledge all of them were
dedicated to study a numerous variety of the linear diffusion equations. Being motivated by
this fact we have decided to explore probably not yet studied effect of stochastic resetting
on the non-linear diffusion. In the present paper we have restricted our attention exclusively
to the simplest, canonical version of the exponential (Poissonian) resetting. Thereby, we save
a detailed analysis of the other schemes of stochastic resetting, interesting from theoretical
and experimental points of view, for the future work.

The summary of the present work is as follows. At the beginning we have invoked a specific
form of the non-linear differential equation in which the diffusion coefficient depends on the
probability density through the power-law function. Next, we have briefly described the
effective method of how to solve this type of a differential equation and then commented its
general solution. Getting a minimal experience in this subject has allowed us to examine how
the Poissonian resetting affects the basic properties of the non-linear diffusion. It turned
out that unlike ordinary diffusion the MSD for the non-linear diffusion firstly increases
with time to finally achieve the constant value over much longer time scales. Such a behavior
of the MSD indicates that also the time-dependent PDF should attain under the influence of
stochastic resetting a permanent profile in the steady state. Indeed, we have found such
stationary distributions in the form of two exact results that refer to peculiar values of the
power-law exponent $\sigma\!=\!1$ and $2$, and the approximate formula when the values of the
parameter $\sigma$ fall into the range between $1$ and $2$. In order to obtain this last
expression we have used the algebraic approximation consisting in a replacement of the
integrand in Eq.~(\ref{eq_28}) with the power-law function arising from the Taylor expansion.
We have also tested the second approach based on the exponential approximation and derived
the alternative formula for the stationary PDF. In this case, the PDF depends on the
position of a diffusing particle according to the stretched exponential tail
$p_{r}(x)\!\simeq\!\exp\left(-\eta x^{\frac{2(\sigma+2)}{\sigma+4}}\right)$ for large $x$.
We have demonstrated that if $\sigma\!\to\!0$ the complete form of the PDF boils down to the
two-sided exponential probability density that results from a solution of the linear diffusion
equation extended with additional terms accounting for exponential resets \cite{Evans2011}.
However, in contrast to the PDF for the linear diffusion the spatial domain of the PDF for
the non-linear diffusion is restricted to a confined support outside of which this function
disappears. Our studies have shown that the mechanism of exponential resetting extends this
finite support of the PDF to the entire domain of real numbers.

In this paper, we have also analysed the process of non-linear diffusion with the Poissonian
resetting from a perspective of its first-passage properties. To that end we assumed the
system with a totally absorbing barrier (the target) placed at the origin of the semi-infinite
interval and an initial position of a diffusing particle at a distant point. As the main
result we have derived the exact and approximate expressions for the mean time to absorption
(the mean first-passage time). This quantity has been studied as the function of the resetting
rate and different values of the parameter $\sigma$. The conclusion emerging from these studies
is that the exponential resetting makes the mean time towards the absorbing point to be finite
and even the shortest one for a particular value of the resetting rate. This means the mean
time needed for a diffusing particle to reach a target attains the global minimum for a
given optimal resetting rate $r_{\ast}$. Furthermore, we have shown that $r_{\ast}$ decreases
with the increase of a distance between an initial position $x_{0}$ and the target.
A dependence of $r_{\ast}$ on and the power-law exponent $\sigma$ displays a similar
monotonic decrease. Finally, we have tested and confirmed the universal property that
the relative fluctuation in the mean first-passage time of optimally restarted non-linear
diffusion intermittent by exponential resetting is equal to unity.

The present paper constitutes only a prelude to more advanced investigations of non-linear
processes evolving under a given dynamics disturbed by stochastic resetting. In particular,
it will be interesting to analyse the non-linear diffusion in a presence of external
potentials, under the influence of non-exponential resetting protocols, under
non-instantaneous returns and explore the temporal relaxation of this process to the
stationary state, to name but a few research directions. We are hopeful that the present work
will initiate a new route for theoretical and experimental studies of stochastic resetting
in the context of various non-linear phenomena.

\newpage

\setcounter{equation}{0}
\renewcommand{\theequation}{A.\arabic{equation}}
\section*{Appendix}

The goal of this supplementary section is to present the formal derivation of
Eq.~(\ref{eq_32}) included in the main text. Setting the parameter $\sigma\!=\!2$ we get
from Eqs.~(\ref{eq_7}) and \ref{eq_8} that two numerical factors appearing in
Eq.~(\ref{eq_28}) take the following values, $\mathcal{A}(2)\!=\!\frac{1}{\sqrt{\pi}}$ and
$\mathcal{B}(2)\!=\!\frac{\pi}{4}$. In addition, the auxiliary function in this equation
is $\xi(x,2)\!=\!\left(\frac{\pi}{4}\right)^{2}\frac{(x-x_{0})^{4}}{D}$. Thus, inserting
all these quantities into Eq.~(\ref{eq_28}) we obtain
\begin{equation}
p_{r}(x\!\mid\!x_{0})=\frac{r}{\sqrt{\pi}}\bigintsss_{\xi(x,2)}^{\infty}
\frac{\mathrm{e}^{-rt}}{(Dt)^{\frac{1}{4}}}\sqrt{1-\frac{\pi(x-x_{0})^{2}}
{4\sqrt{Dt}}}\,\mathrm{d}t.
\label{eq_A1}
\end{equation}
Now, upon introducing the new notation $a(x)\!\equiv\!\sqrt{\xi(x,2)}$ and changing the
variable of integration from $t$ to $\tau$, so that $t\!=\!(\tau^{2}+a(x))^{2}$, we can
transform the PDF in Eq.~(\ref{eq_A1}) into the much simpler form:
\begin{equation}
p_{r}(x\!\mid\!x_{0})=\frac{4r}{\sqrt{\pi\sqrt{D}}}\int_{0}^{\infty}\tau^{2}\exp\left[
-r(\tau^{2}+a(x))^{2}\right]\,\mathrm{d}\tau.
\label{eq_A2}
\end{equation}
From now on the rest of our calculations boils down to the solution of the above integral.
It can be find by using the similarly looking integral \cite{Gradshteyn2007}
\begin{equation}
\int_{0}^{\infty}\exp\left(-\alpha\tau^{4}-2\beta\tau^{2}\right)\mathrm{d}\tau=
\frac{1}{2}\sqrt{\frac{\beta}{2\alpha}}\exp\left(\frac{\beta^{2}}{2\alpha}\right)
\mathrm{K}_{\frac{1}{4}}\left(\frac{\beta^{2}}{2\alpha}\right),
\label{eq_A3}
\end{equation}
where the both factors $\alpha$, $\beta\!>\!0$ and $\mathrm{K}_{\nu}(z)$ is the modified
Bessel function of the second kind. Out of many well-known properties of this function
we utilize the two ones:
\begin{equation}
\mathrm{K}_{\nu}(z)=\mathrm{K}_{-\nu}(z),
\label{eq_A4}
\end{equation}
and
\begin{equation}
\frac{\mathrm{d}}{\mathrm{d}z}\mathrm{K}_{\nu}(z)=-\mathrm{K}_{\nu-1}(z)-\frac{\nu}{z}
\mathrm{K}_{\nu}(z).
\label{eq_A5}
\end{equation}
At first, however, let us differentiate both sides of Eq.~(\ref{eq_A3}) with respect to
the parameter $\beta$. In this way, we have
\begin{equation}
\int_{0}^{\infty}\tau^{2}\exp\left(-\alpha\tau^{4}-2\beta\tau^{2}\right)\mathrm{d}\tau=
-\frac{1}{4\sqrt{2\alpha}}\frac{\mathrm{d}}{\mathrm{d}\beta}\left[\sqrt{\beta}
\exp\left(\frac{\beta^{2}}{2\alpha}\right)\mathrm{K}_{\frac{1}{4}}\left(\frac{\beta^{2}}
{2\alpha}\right)\right].
\label{eq_A6}
\end{equation}
Taking into account Eqs.~(\ref{eq_A4}) and (\ref{eq_A5}) we show that a derivative of
the expression enclosed in the square bracket on the right hand side of the above equation
\begin{equation}
\frac{\mathrm{d}}{\mathrm{d}\beta}\left[\sqrt{\beta}
\exp\left(\frac{\beta^{2}}{2\alpha}\right)\mathrm{K}_{\frac{1}{4}}\left(\frac{\beta^{2}}
{2\alpha}\right)\right]=\frac{\beta\sqrt{\beta}}{\alpha}\exp\left(\frac{\beta^{2}}{2\alpha}
\right)\left[\mathrm{K}_{\frac{1}{4}}\left(\frac{\beta^{2}}{2\alpha}\right)-
\mathrm{K}_{\frac{3}{4}}\left(\frac{\beta^{2}}{2\alpha}\right)\right].
\label{eq_A7}
\end{equation}
If we plug this derivative back into Eq.~(\ref{eq_A6}) and multiply their both sides by
the exponential function $\exp(-\frac{\beta^{2}}{\alpha})$, we obtain that the integral
\begin{equation}
\bigintsss_{0}^{\infty}\tau^{2}\exp\left[-\alpha\left(\tau^{2}+\frac{\beta}{\alpha}\right)^{2}
\right]\mathrm{d}\tau=\frac{1}{4\sqrt{2}}\left(\frac{\beta}{\alpha}\right)^{\frac{3}{2}}
\exp\left(-\frac{\beta^{2}}{2\alpha}\right)\left[\mathrm{K}_{\frac{3}{4}}\left(
\frac{\beta^{2}}{2\alpha}\right)-\mathrm{K}_{\frac{1}{4}}\left(\frac{\beta^{2}}{2\alpha}
\right)\right].
\label{eq_A8}
\end{equation}
Having this equation at our disposal and taking $\alpha\!=\!r$ and $\beta\!=\!ra(x)$,
we immediately find out the integral in Eq.~(\ref{eq_A2}) and hence the exact formula
expressing the PDF with the parameter $\sigma\!=\!2$ for the non-linear diffusion
intermittent by exponential resets. The final result is as follows
\begin{equation}
p_{r}(x\!\mid\!x_{0})=\frac{ra^{\frac{3}{2}}(x)}{\sqrt{2\pi\sqrt{D}}}\exp\!
\left(-\frac{ra^{2}(x)}{2}\right)\left[\mathrm{K}_{\frac{3}{4}}\!\left(\frac{ra^{2}(x)}{2}
\right)-\mathrm{K}_{\frac{1}{4}}\!\left(\frac{ra^{2}(x)}{2}\right)\right],
\label{eq_A9}
\end{equation}
where the variable $a(x)\!=\!\frac{\pi(x-x_{0})^{2}}{4\sqrt{D}}$.

To complete this supplement we should also determine the value of the PDF for $x\!=\!x_{0}$.
It can be done in two ways. The first method takes advantage of the following limit values of
the modified Bessel functions of the second kind:
\begin{equation}
\lim_{z\to0}z^{3}\mathrm{K}_{\frac{1}{4}}\left(uz^{4}\right)=0,
\label{eq_A10}
\end{equation}
and
\begin{equation}
\lim_{z\to0}z^{3}\mathrm{K}_{\frac{3}{4}}\left(uz^{4}\right)=
\frac{\Gamma\left(\frac{3}{4}\right)}{\sqrt[4]{2}\,u^{3/4}}.
\label{eq_A11}
\end{equation}
Calculating these limits in Eq.~(\ref{eq_A9}) we easily show that
\begin{equation}
p_{r}(x_{0})=\frac{\Gamma\left(\frac{3}{4}\right)}{\sqrt{\pi}}\left(\frac{r}{D}\right)
^{\frac{1}{4}}.
\label{eq_A12}
\end{equation}
The second method that guarantees the above result consists in the direct calculation of
the integral in Eq.~(\ref{eq_A2}) for $x\!=\!x_{0}$ with $\xi(x_{0},2)\!=\!0$. In this case
we use the integral representation of the Euler gamma function, i.e.
$\Gamma(\alpha)\!=\!\int_{0}^{\infty}u^{\alpha-1}\mathrm{e}^{-u}\mathrm{d}u$.

\section*{Conflicts of Interest}
The author declares no conflict of interest.

%Bibliography
\bibliographystyle{unsrt}  
\bibliography{paper}  

\end{document}